\documentclass[english,12pt]{emulateapj}
\usepackage{rotating}
\usepackage[T1]{fontenc}
\usepackage[latin1]{inputenc}
\usepackage{verbatim}
\usepackage{graphicx}
\usepackage{amssymb}
\usepackage{color}
\makeatletter

\usepackage{babel}
\makeatother

\begin{document}

\title{Unstable Planetary Systems Emerging Out Of Gas Disks}

\author{Soko Matsumura\altaffilmark{1},
Edward W.~Thommes\altaffilmark{1,2},
Sourav Chatterjee\altaffilmark{1}}
\author{Frederic A.~Rasio\altaffilmark{1}}
%
%
%
%
%

\altaffiltext{1}{Department of Physics and Astronomy, Northwestern
University, Evanston, IL 60208, USA}

\altaffiltext{2}{University of Guelph, Guelph, ON, Canada}

\begin{abstract}
The discovery of over 400 extrasolar planets allows us to statistically test our
understanding of formation and dynamics of planetary systems via numerical simulations. 
Traditional N-body simulations of multiple-planet systems without gas disks have successfully reproduced the
eccentricity ($e$) distribution of the observed systems, by assuming
that the planetary systems are relatively closely packed when the gas disk
dissipates, so that they become dynamically unstable within the
stellar lifetime. However, such studies cannot explain the small
semimajor axes $a$ of extrasolar planetary systems, if planets are
formed, as the standard planet formation theory suggests, beyond the ice line.

In this paper, we numerically study the evolution of three-planet systems in dissipating gas disks, and 
constrain the initial conditions that reproduce the observed $a$ and $e$ distributions simultaneously. 
We adopt the initial conditions that are motivated by the standard planet formation theory, and self-consistently simulate 
the disk evolution, and planet migration by using a hybrid N-body and 1D gas disk code. 
We also take account of eccentricity damping, and investigate the effect of saturation of corotation resonances on the evolution of planetary systems.
We find that the $a$ distribution is largely determined in a gas disk, while the $e$ distribution is determined after the disk dissipation. 
We also find that there may be an optimum disk mass which leads to the observed $a-e$ distribution. 
Our simulations generate a larger fraction of planetary systems trapped in mean-motion resonances (MMRs) than the observations, 
indicating that the disk's perturbation to the planetary orbits may be important to explain the observed rate of MMRs.  
We also find much lower occurrence of planets on retrograde orbits than the current observations of close-in planets suggest. 
\end{abstract}
\keywords{methods: numerical, n-body simulations, planetary systems:
protoplanetary disks, formation, planets and satellites: formation,
general}

\section{Introduction}

Out of over 360 planetary systems discovered so far, about 12.4\%
are known to be multiplanet systems (http://exoplanet.eu/). 
Also, recent observations have started revealing that many of the detected
planets are accompanied by a planet on a further orbit
\citep[e.g.][]{Wittenmyer07,Wright07}. It will become increasingly
more important to understand the formation and evolution of multiplanet systems, which can explain the observed properties of 
extrasolar planetary systems.

Recent numerical N-body simulations of planetary systems
\emph{without} a gas disk demonstrated that dynamical instabilities
occurring in the multiplanet systems, which are characterized by
orbital crossings, collisions, and ejections of planets, could
increase planetary eccentricities ($e$) efficiently
\citep[e.g.,][]{Rasio96,Weidenschilling96}. These studies successfully
reproduced the observed eccentricity distribution of extrasolar
planets \citep[][from here on C08, and JT08,
respectively.]{Ford08,Chatterjee08,Juric08}

Such N-body simulations also suggest that the planet--planet
interactions alone cannot explain small semimajor axes ($a$) of the
observed planets, \emph{if} giant planets are formed beyond the ice
line as expected from the standard planet formation theory. 
More specifically, starting with giant planet systems beyond $3\,$AU, C08 found that 
it is difficult to scatter planets within $\sim1\,$AU.
This is because planet--planet interactions are not
particularly efficient in shrinking the planetary orbits.

The disk--planet interactions, on the other hand, are known to
decrease semi-major axes of planets efficiently \citep{Ward97}. 
The overall effect of such interactions on the orbital eccentricity is highly uncertain, and depends on a 
detailed disk structure, as well as planetary masses.
Generally, disk--planet interactions lead to eccentricity damping, but for planets massive enough to open a clean gap in the disk, 
eccentricity can increase rapidly depending on the level of saturation of corotation resonances \citep{Goldreich03,Moorhead08}. 
However, hydrodynamic simulations show that the disk--planet interactions typically lead to $e<0.2$ \citep[e.g.,][]{DAngelo06}.

The time to dynamical instability scales with the distance between planets (e.g.,~C08). 
Thus, all the N-body studies on planet--planet scattering assume initially dynamically active planetary systems 
(i.e.,~distance between planets being less than about a few Hill radii).   
However, with the aid of a gas disk, planets which would not easily reach dynamical instability could experience strong interactions. 
For example, when planets are embedded in the inner cavity of a disk, the surrounding disk would push the outer planet 
closer to the inner one, triggering the dynamical instability \citep[e.g.,][]{Adams03,Moorhead05}.  
The orbital eccentricity can also be increased if planets are trapped in mean motion resonances (MMRs) 
during such a convergent migration \citep[e.g.,][]{Snellgrove01,Lee02}.  
Alternatively, when a gas disk annulus is left between planets, the eccentricities of planets could 
still increase by repeated resonance crossings due to divergent migration \citep{Chiang02}.
Thus, whether the combined effects of disk--planet and planet--planet interactions would lead to 
eccentricity excitation, or damping should be studied carefully. 
One of the goals of our study is to verify the initial assumptions of N-body studies (i.e.,~planets are on nearly circular, and coplanar orbits 
when the gas disk is around), and figure out whether the eccentricity distribution is largely determined before, or after the disk dissipation. 

In this paper, we numerically study the evolution of three-planet systems 
in a dissipating gas disk, and constrain the ``initial'' conditions of 
planetary systems which can reproduce the $a$ and $e$ distributions simultaneously.  
We calculate the disk--planet interactions directly so that the disk and planetary orbits evolve self-consistently.  Also, we take account of the effect of saturation of corotation resonances on eccentricity damping.
We introduce the numerical methods in Section 2, and the initial conditions in Section 3. 
In Section 4, we show that the observed $a-e$ scattered plot can be reproduced well for 
a reasonable range of disk masses.  We also discuss the mass distribution, and mean-motion resonances for representative cases.
Finally in Section 5, we compare this work with some recent observations, and summarize our results.

\section{Numerical Methods and Initial conditions}

To simulate multiplanet systems in gas disks, we use a hybrid code
which combines the symplectic N-body integrator SyMBA \citep{Duncan98}
with a one-dimensional gas disk evolution code \citep{Thommes05}.
SyMBA utilizes a variant of the so-called mixed-variable symplectic
(MVS) method \citep{Wisdom91}, which treats the interaction between
planets as a perturbation to the Keplerian motion around the central
star, and handles close encounters between bodies with a force-splitting
scheme which has the properties of an adaptive timestep \citep{Duncan98}.
When the bodies are
well-separated, SyMBA has the speed of the MVS method, while during
the close encounters, the timestep for the relevant bodies is
recursively subdivided.

On the other hand, the gaseous disk evolves both viscously and via
gravitational interaction with planets, according to a general
Navier-Stokes equation. Following the standard prescription by
\citet{Lin86a}, the gas disk is divided into radial bins, which
represent disk annuli with azimuthally and vertically averaged disk
properties like surface mass density, temperature, and viscosity.
Viscous evolution of the disk is calculated by using the standard
alpha viscosity prescription \citep{Shakura73}, while the
disk--planet interactions modify the disk evolution via the torque
density formulated in \citet{Ward97} \citep[see also][]{Menou04}.
The calculated torque density is used in turn to determine the
migration rates of planets.

In our simulations, a disk stretches from $0.02$ to $100$ AU, and
the orbital evolution of planets is followed down to $0.02$ AU.  The
timestep of simulations is typically $0.05$ yr, which ensures a
reasonable orbital resolution down to $\sim 0.2$ AU.

\subsection{Gas Accretion onto a Planet}
The above code follows the evolution of planetary
systems as planets gravitationally interact with each other, migrate,
and open gaps in the disk. However, planets are also expected to
clear the gas annuli between them as they grow by accreting gas from
the surrounding disk. 

Once a planet becomes massive enough to open a gap in the disk, the
gas accretion rate is controlled by how quickly the disk can supply
gas to the planet, rather than how quickly the planet can accrete
gas \citep{Bryden00,Tanigawa02}. Since all the planets in our
simulations are more massive than the Neptune, they are expected to
have circumplanetary disks, from which they accrete. Although our
code does not resolve such disks, we can mimic the accretion effect
by adopting the results of hydro simulations. \cite{Tanigawa02} showed that,
without a gap-opening effect, a planet accretes gas within a few Hill radii
on the following timescale.
\begin{equation}
\tau_{subdisk}=30.76~{\rm yr}~\left(\frac{M_p}{M_J}\right)^{-1/3}
\left(\frac{\Sigma(r)}{1.7\times10^3~{\rm g\,cm^{-2}}}\right)^{-1}
\label{tau3}
\end{equation}
%
Here, $M$ is mass, and $\Sigma$ is the surface mass density. 
The subscripts $p$, and $J$ represent the planet, and Jupiter, respectively.
Due to the gap-opening effect, the true accretion timescale is
likely longer than the above estimate \citep[e.g.][]{DAngelo03}.
However, since our code handles a gap-opening due to disk-planet
interactions separately, we adopt the above accretion timescale for all of our
planets.

\subsection{Eccentricity Evolution}
For simplicity, we focus on eccentricity evolution due to the first-order resonances.
Also, to evaluate the effects of eccentricity excitation due to planet--planet interactions,
we neglect the $e$ excitation due to disk--planet interactions.

When a planet is too small to open clean gaps, the eccentricity
evolution is dominated by co-orbital Lindblad resonances which damp
eccentricity \citep[e.g.][]{Artymowicz93a}. In such a case, the
eccentricity damping timescale can be expressed as follows
\citep[e.g.][]{Kominami04}.
\begin{equation}
\tau_{edamp}=-\frac{e_{p}}{\dot{e_{p}}}=\left(\frac{h_p}{r_p}\right)^4
\left(\frac{M_{*}^{2}}{M_{p}\bar{\Sigma_p} r_p^{2}}\right)\Omega(r_p)^{-1}
\label{eq0}
\end{equation}
Here, $h$ is the pressure scale height, $r$ is the orbital radius, and $\Omega$ is the orbital frequency.
The subscript $*$ denotes the star.
Note that the equation is evaluated at the location of a
planet. Also, for the disk's surface mass density, we take the
average of peri-center, semi-major axis, and apo-center of a
planetary orbit.

On the other hand, when a planet becomes large enough to open a
clean gap, the effect of co-orbital Lindblad resonances diminishes,
and the competing effects of $e$ excitation due to external Lindblad
resonances and $e$ damping due to corotation resonances become
significant \citep{Goldreich80}. In such a case, the eccentricity
damping can be written as follows \citep{Goldreich03}:
\begin{equation}
\tau_{edamp}=-\frac{e_{p}}{\dot{e_{p}}}=
\frac{1}{K_{e}}\left(\frac{w}{r}\right)^4
\left(\frac{M_{*}^{2}}{M_{p}\Sigma r^{2}}\right)\Omega^{-1} \,
\label{eq1}
\end{equation}
where $w=r(3\pi\alpha)^{-1/3}(r/h)^{2/3}(M_p/M_*)^{2/3}$ is a gap
width which is determined by balancing the gap-opening tidal torque
from the principal Lindblad resonances with the gap-closing viscous
torque, and $\alpha$ is the standard viscosity parameter by \cite{Shakura73}. The above equation is evaluated at the gap edges, where the
contribution from the resonances is the largest. The damping
efficiency is governed by $K_{e}$, which is defined as below by
following the approach of \citet{Goldreich03}.
%
\begin{equation}
K_{e}=\left[1.046\, F(p) - 1 \right]. \label{eq2}
\end{equation}
In this equation, $F(p)$ is the saturation function of corotation
torques that is numerically evaluated by \citet{Ogilvie02} and
interpolated by \citet{Goldreich03} as
\begin{eqnarray}
F(p) & \simeq &
\frac{\left(1+0.65\, p^{3}\right)^{5/6}}{(1+1.022\, p^{2})^{2}}\label{eq3}\\
p & \sim &
\left(\frac{r}{h}\right)^{2/9}(\frac{M_{*}}{M_{p}})^{1/9}\frac{e_{p}}{\alpha^{1/9}}
. \label{eq4}
\end{eqnarray}
%
For $K_{e}=0.046$ (and hence $F(p)=1$), the corotation torques are
unsaturated, and fully contribute to the eccentricity damping, while
for the other extreme $K_{e}=0$, the effect of corotation torques
are negligible, and there is no damping. Negative values of $K_{e}$
corresponds to $e$ excitation, but we don't take account of the
effect here.
Note, however, that hydrodynamic simulations show that the disk--planet
interactions typically lead to $e<0.2$ \citep[e.g.][]{DAngelo06},
and therefore may not be able to explain planets with high
eccentricities.

\subsection{Disk Dissipation Timescale}
C08 suggested that the dynamical instability occurs more frequently
as the disk mass decreases. Generally, the lifetime of gas disks is
estimated to be $1-10$ Myr
\citep[e.g.][]{Hillenbrand05ap,SiciliaAguilar06}, but the mechanism
of the final dispersal of disks is not well-understood.

Observations suggest that such a timescale is rather short, $\sim
10^5-10^6$ yr \citep[e.g.][]{Simon95,Currie09}.  Since the viscous accretion
timescale of a disk is longer than this, currently the most
promising mechanism to explain the rapid dispersal of disks is
photoevaporation, which can remove a disk within $10^5-10^6$ yr in
favorable cases \citep{Matsuyama03,Alexander06b}. Here, we simply
treat the gas disk dissipation time as a parameter, and assume that
the entire disk is removed exponentially once $\tau_{GD}$ is
reached.  In a Jupiter-mass disk, this disk removal timescale is about $10^5$ yr.   

\section{Initial Conditions}
We study evolution of three-planet systems in a gas disk by changing
various parameters. Specifically, we study five different disk
masses with $200$ planetary systems, by changing gas dissipation
time $\tau_{GD}$ between $2-4\,$Myr. We run each set of runs with and without
the effect of saturation of corotation resonances. Our initial
conditions are summarized in Table~\ref{tab1}.

We focus on three-planet systems, and define their initial
properties following C08. 
We determine the planetary masses by adopting a simplified core accretion scenario.
Specifically, we sample planetary core masses randomly from a distribution over 
$1-100\,M_E$  (where $M_{E}$ is the Earth mass) uniform in $M_{core}^{1/5}$, and 
assume that each core accretes all gas within the ``feeding zone'' that is extending over $\Delta=8R_{hill,\, core}$, and
centered on the core:
\[
M_{p}=2\pi a\Delta\Sigma+M_{core}.\]
Here, the core's Hill radius is defined as $R_{hill,\,
core}=(M_{core}/(3M_{*}))^{1/3}a$. The size of the feeding zones is
a typical distance between planetary
embryos \citep{Kokubo02}. 

As in C08, the semimajor axes are chosen so that the distance
between planets is scaled with $K=4.4$ times the Hill radius of the
i-th planet: \[ a_{i+1}-a_{i}=K\, R_{hill,i},\] with $a_{1}=3$AU. We
fix the initial semimajor axis of the innermost planet following the
common assumption that giant planets form beyond the ``ice line'',
where the solid density is higher due to the condensation of icy
and/or carbonaceous material \citep{Lewis74,Lodders04}.
From this prescription, we obtain planets with mass ranging over
$0.4-4\, M_{J}$, between $3$ to $6.5$ AU. 

We make two independent sets of 100 three-planet systems. Since
typical planetary eccentricities and inclinations just after planet
formation are not known, we use two different ranges of initial $e$
and $i$.
For one of the sets (marked with $d1$ in Table~\ref{tab1}), we
choose initially moderate $e$ and $i$, which are randomly drawn from
a uniform distribution in the range of $e=0-0.1$, and from a
distribution uniform in $\cos i$ over the range of $i=0-10\,$deg,
respectively. For the latter set (marked with $d2$ in
Table~\ref{tab1}), on the other hand, we choose initially small $e$
and $i$ over the range of $e=0-0.05$, and $i=0-0.03\,$deg,
respectively. The former set is identical to \emph{Mass distribution
3} in C08.
In both sets, phase angles are randomly chosen from a uniform
distribution over $0-360\,$deg.  Also, gas disk dissipation time $\tau_{GD}$ is 
chosen randomly between $2-4$ Myr for each three-planet system.
%

Since our planets are nearly fully grown, we use evolved gas disks as the initial gas disks, 
instead of using the primordial ones.
Each initial disk is generated by evolving the minimum mass solar
nebula (MMSN)-type disk with $\Sigma=10^{3}(a/AU)^{-3/2}\, g\,
cm^{-2}$ for 5, 6, 7, 8, and 9 Myr without planets, under the disk's
viscosity $\alpha=5\times10^{-3}$.
This corresponds to five different initial disk masses in the range of
$\sim 0.3-1.6 M_J$. All disks are stretching from $0.02$ to $100\,$AU.

These initial conditions are summarized in Table~\ref{tab1}.
%
%
\begin{deluxetable}{ccccc}
\tablecolumns{5} \small \tablewidth{0pt} \tablecaption{Initial Disk
Conditions \label{tab1}}
\tablehead{ \colhead{Set No.} & \colhead{$\tau_{disk}$[Myr]} &
\colhead{$M_{disk}$[$M_{J}$]} & \colhead{Distribution}  &
\colhead{$K_{e}$} } 
\startdata
t5d1 & 5 & 1.6 & 1 & 0.046 \\
t6d1 & 6 & 1.1 & 1 & 0.046 \\
t7d1 & 7 & 0.71 & 1 & 0.046 \\
t8d1 & 8 & 0.47 & 1 & 0.046 \\
t9d1 & 9 & 0.32 & 1 & 0.046 \\
t5d1cr & 5 & 1.6 & 1 & CR \\
t6d1cr & 6 & 1.1 & 1 & CR \\
t7d1cr & 7 & 0.71 & 1 & CR \\
t8d1cr & 8 & 0.47 & 1 & CR \\
t9d1cr & 9 & 0.32 & 1 & CR \\
t5d2 & 5 & 1.6 & 2 & 0.046 \\
t6d2 & 6 & 1.1 & 2 & 0.046 \\
t7d2 & 7 & 0.71 & 2 & 0.046 \\
t8d2 & 8 & 0.47 & 2 & 0.046 \\
t9d2 & 9 & 0.32 & 2 & 0.046 \\
t5d2cr & 5 & 1.6 & 2 & CR \\
t6d2cr & 6 & 1.1 & 2 & CR \\
t7d2cr & 7 & 0.71 & 2 & CR \\
t8d2cr & 8 & 0.47 & 2 & CR \\
t9d2cr & 9 & 0.32 & 2 & CR \\
%
\enddata
\tablecomments{Initial conditions for each set of 100 runs. Column 1
shows names of sets, where $t(5-9)$ indicates the age of initial
disks, $d1$ and $d2$ represent two different distributions of 100
three-planet systems ($d1$ has larger initial $e$ and $i$ compared
to $d2$, see Section 3), and $cr$ for sets of runs taking account of
saturation of corotation resonances. Column 2 shows the age of
initial disks, which is obtained by evolving a MMSN-type disk under
$\alpha=5\times10^{-3}$. Column 3 lists the corresponding initial
disk mass.
Column 4 shows two different distributions of three-planet systems.
Column 5 shows the eccentricity damping factor $K_{e}$, where $0.046$ is
full damping, and $CR$ means that the saturation of corotation
resonances is taken into account (see Section 2.2).}
\end{deluxetable}
%

\section{Results}
%
In Sections 4.1 and 4.2, we focus on the results of sets with
initially moderate $e$ and $i$ (i.e.,~sets denoted by $d1$, see
Table~\ref{tab1}). Without a gas disk, C08 showed that N-body
simulations of this type of three-planet systems reproduce the
observed $e$ distribution of exoplanets very well. 
On the other hand, by assuming that planets are formed beyond the ice-line ($\sim3\,$AU), 
their simulations showed that it is difficult to scatter planets within $\sim1\,$AU.
In Section 4.1, we discuss the effects of the initial disk mass on the $a-e$
distribution, while in Section 4.2, we study the effects of the
saturation of corotation resonances. In Section 4.3, we investigate
the corresponding results for initially small $e$ and $i$.  We
discuss typical evolution cases in Section 4.4, and mass
distribution in Section 4.5. Finally in Section 4.6, we study 
the mean motion resonances seen in our simulations.

\subsection{Effect of the initial disk mass}
\begin{figure}
\plotone{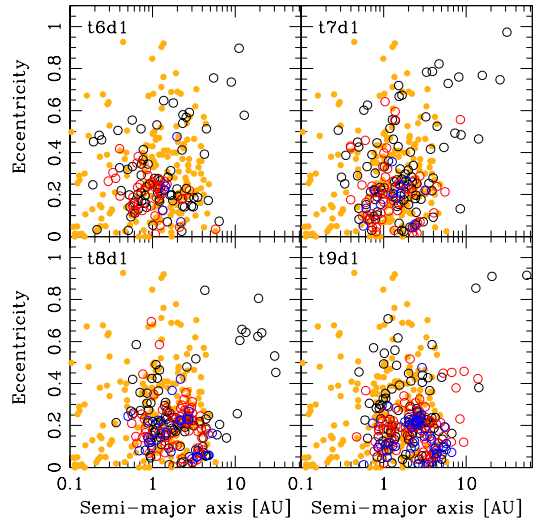}

\caption{The final $a-e$ scattered plot for $t6d1-t9d1$. Black, red,
and blue circles indicate the innermost, middle, and outermost
planet, respectively. Orange solid circles are the observed exosolar
planets. The 2D K-S tests for the observed and simulated
distributions show that the null hypothesis cannot be rejected for
$t6d1$ and $t7d1$. \label{fig1}}
\end{figure}

First, we investigate the effects of initial disk masses on final
distribution of $a$ and $e$ by neglecting the effects of saturation
of corotation resonances (i.e.,~$t5d1-t9d1$).  The $a-e$ scattered
plots of planets after $100\,$Myr for $t6d1-t9d1$ are plotted in
Fig.~\ref{fig1}. Also plotted is the observed $a-e$ distribution. 
As expected, more massive disks have more planets with
smaller semi-major axes, due to efficient planet migration. 
%

The overall trend of the $a-e$ scattered plot is reproduced 
fairly well, especially for $t6d1$ and $t7d1$.  We can quantify
this by using the Kolmogorov-Smirnov (K-S) test. We perform the K-S
tests against the null hypothesis that two arbitrary distributions
are drawn from the same underlying distribution, and quote the
significance level probabilities in Table~\ref{tab3}. We choose to
reject the null hypothesis for $P<0.1$. To compare the observed and
simulated distributions, we select planets between $0.2$ and $6$ AU with mass ranging over $0.3-4 M_J$. 
The lower limit of semi-major axis comes from the
resolution limit of our simulations (see Section 2), while the upper
limit is motivated by the current maximum orbital radius of a planet
detected by radial-velocity observations.  Our two dimensional K-S
tests indeed show that we cannot reject the null hypothesis of the observed
and simulated distributions for $t6d1$ and $t7d1$.
The agreement with the observed distribution becomes worse for more, or 
less massive disk cases ($t5d1$, $t8d1$, and $t9d1$).
Thus, our results indicate that there may be an optimum disk mass to reproduce the observed $a-e$ distribution.

\begin{figure}
\plotone{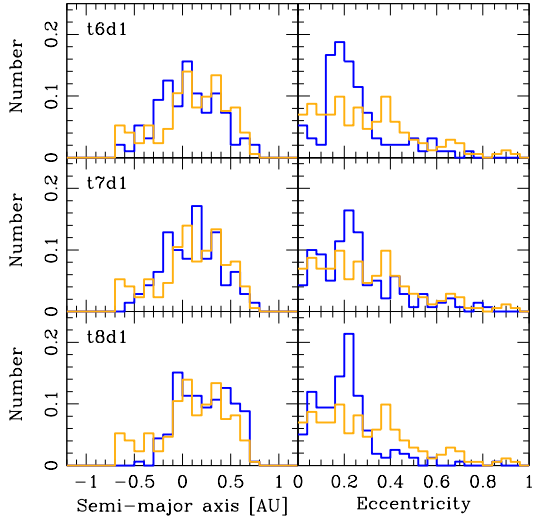}

\caption{Left: Final $a$ distributions for $t6d1$, $t7d1$, and
$t8d1$ (blue histograms), compared with the observed distribution
(orange histograms). Right: Corresponding $e$ distributions. From
the 1D K-S tests, the null hypothesis of the observed and simulated
$a$ distributions cannot be rejected for $t6d1$ and $t7d1$.
Similarly, the null hypothesis for the $e$ distributions cannot be
rejected for $t7d1$. \label{fig2}}
\end{figure}

Now we look into the $a$ and $e$ distributions separately.
Fig.~\ref{fig2} compares the observed and simulated $a$ and $e$
distributions at $100\,$Myr for $t6d1$, $t7d1$, and $t8d1$. For $a$
distribution, the K-S test shows that we cannot reject the null
hypothesis for $t6d1$ and $t7d1$ at $100\,$Myr, while for $t8d1$, the observed and 
simulated semi-major axis distributions are significantly different from each other.  
Thus, our results indicate that a disk mass less than $\sim0.5 M_J$ does not lead to significant planet
migration.  Efficient planet migration requires that an outer disk
mass is comparable to, or larger than, the planetary mass
\citep{Ivanov99}. Since the mass range of our planets is $0.4-4 M_J$, it is expected 
that lower disk mass cases have little planetary
migration.

For $e$ distribution, on the other hand, the K-S test shows that the
null hypothesis cannot be rejected for $t7d1$ at $100\,$Myr. The
distributions of $t8d1$ and $t9d1$ are dominated by planets with
relatively small eccentricities ($e\lesssim0.3$), indicating that 
these cases are more dynamically stable compared to $t7d1$. 
On the other hand, $t6d1$ has a deficit of planets with $e\lesssim0.1$ and an
overabundance of planets with $e\sim0.2$. 
A similar deficit of planets with low eccentricities is reported by JT08 for disk-less cases.
%

Now, we compare the corresponding $a-e$ distributions at the disk dissipation time $\tau_{GD}$ with the observed distribution, 
to highlight the evolution of planetary systems while a gas disk is around.
The 2D K-S tests reject the null hypothesis with the observed
distribution for all of these sets ($t5d1-t9d1$) at $\tau_{GD}$.
This results indicate that planet--planet interactions \emph{after} the gas disk
dissipation must have played a significant role in determining the
final $a-e$ distribution. The 1D K-S tests support this statement.
For the $a$ distribution, the null hypothesis is rejected for all
the sets but $t7d1$ at $\tau_{GD}$, while for the $e$ distribution,
the null hypothesis is rejected for all five sets at $\tau_{GD}$.

Overall, our results imply that the $a$ distribution is largely
determined while a gas disk is around, since the semi-major axis distribution does not 
change dramatically between $\tau_{GD}$ and 100\,Myr. 
The $e$ distribution is largely determined by planet--planet interactions after the disk dissipates, 
because the orbital eccentricities generally stay low while a gas disk is around.
Thus, we find that the initial assumption of nearly circular orbits in the previous N-body studies is reasonable. 
%

\subsection{Effect of the saturation of corotation resonances}
Next, we repeat the same set of simulations by taking account of the
saturation of corotation resonances (i.e.,~$t5d1cr-t9d1cr$). The
overall results turn out to be very similar to $t5d1-t9d1$. In fact,
as it can be seen from Table~\ref{tab3}, the 2D K-S tests against
the observed and final $a-e$ scattered plots show that the null
hypothesis cannot be rejected for $t6d1cr$, and $t7d1cr$. 
The 1D K-S tests for $a$ and $e$ support this result. Furthermore, as in Table~\ref{tab2},
the rates of ejections, collisions, and mergers are comparable for
the cases with and without the saturation of corotation resonances,
both before and after $\tau_{GD}$.  Since we don't take account of
the $e$ excitation due to the disk--planet interactions, the
inclusion of the saturation of corotation resonances only prolongs
the $e$ damping time in our simulations. The fact that such effects
do not dramatically change the final $a-e$ distribution implies that 
the eccentricity damping without the saturation of corotation resonances 
is as inefficient as that with saturation for the range of disk masses we use.  
%

\subsection{Effect of initial $e$ and $i$}
We also study the evolution of three-planet systems with initially
small $e$ and $i$ (sets denoted with $d2$, see Table~\ref{tab1}).
The overall results are similar to the cases with moderate $e$ and
$i$, for both with and without the saturation of corotation
resonances.

For cases \emph{without} the saturation of corotation resonances
($t5d2-t9d2$), the 2D K-S tests against the observed and final
distributions show that the null hypothesis cannot be rejected for
$t6d2$ and $t7d2$.  On the other hand, for cases \emph{with} the saturation
of corotation resonances ($t5d2cr-t9d2cr$), the corresponding tests
show that the null hypothesis cannot be rejected for $t7d2cr$ and
$t8d2cr$.  Thus, again, there seems to be an optimum disk mass to reproduce the 
$a-e$ distribution.

One significant difference for having initially small or moderate $e$
and $i$ values is the rate of mergers. When initial systems are closer to
coplanar, the merger rates tend to be higher. By comparing
$t5d1-t9d1$ with $t5d2-t9d2$, we find that the merger rates before
$\tau_{GD}$ are $\sim1.5-3$ times higher in $t5d2-t9d2$, while the
merger rates after $\tau_{GD}$ are comparable. Similarly, the
comparison of $t5d1cr-t9d1cr$ with $t5d2cr-t9d2cr$ shows that the
merger rates before $\tau_{GD}$ are $\sim1.5-5$ times higher in the
latter sets. We find that most of these mergers occur immediately
after the start of the simulations, within $10^4$ yr. \cite{Ford01}
studied the dynamical evolution of equal-mass two-planet systems,
and found that mergers tend to produce low eccentricity ($e<0.1$)
planets. However, in our simulations, we do not find any
statistically significant change in the fraction of low $e$ planets
between $d1$ and $d2$ cases.

\begin{figure}
\plotone{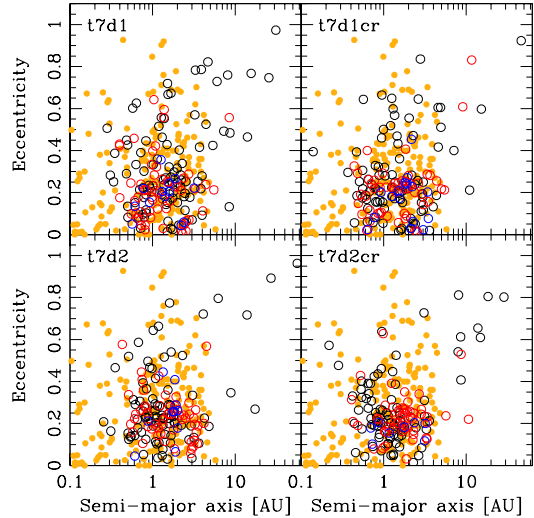}

\caption{The final $a-e$ scattered plot for t7 cases. Upper panels
have moderate initial $e$ and $i$, while bottom panels have small
initial $e$ and $i$. Left panels don't take account of the effect of
saturation of corotation resonances, while right panels do.
\label{fig3}}
\end{figure}

Fig.~\ref{fig3} summarizes the $a-e$ scattered plots for $t7d1$,
$t7d1cr$, $t7d2$, and $t7d2cr$ at $100\,$Myr.  In all cases, the
null hypothesis for the observed and simulated distributions cannot
be rejected for $0.2\,$AU~$\le a \le 6\,$AU.

\subsection{Various Evolution Cases}
Table~\ref{tab2} summarizes the dynamical outcomes of our
simulations, and lists the rates of ejection out of the system,
collision with the central star, and merger between planets, both
before and after $\tau_{GD}$. All the rates become higher for more
massive disks, indicating that stronger convergent/divergent
migration in the disk leads to more frequent dynamical
instabilities. As in the last subsection, the merger rates are
higher for initially small $e$ and $i$ systems (i.e.,~$d2$ systems).

\begin{figure}
\plottwo{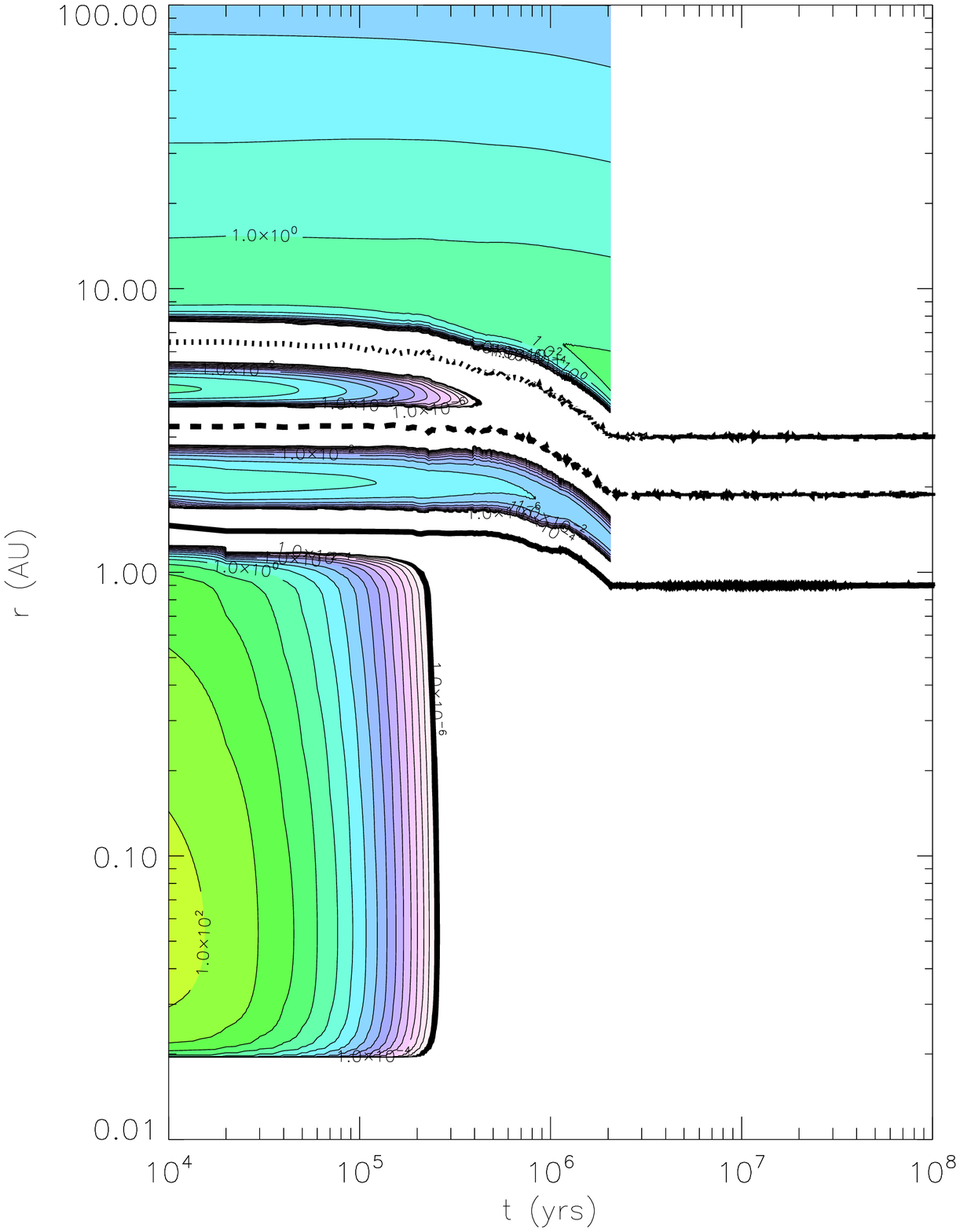}{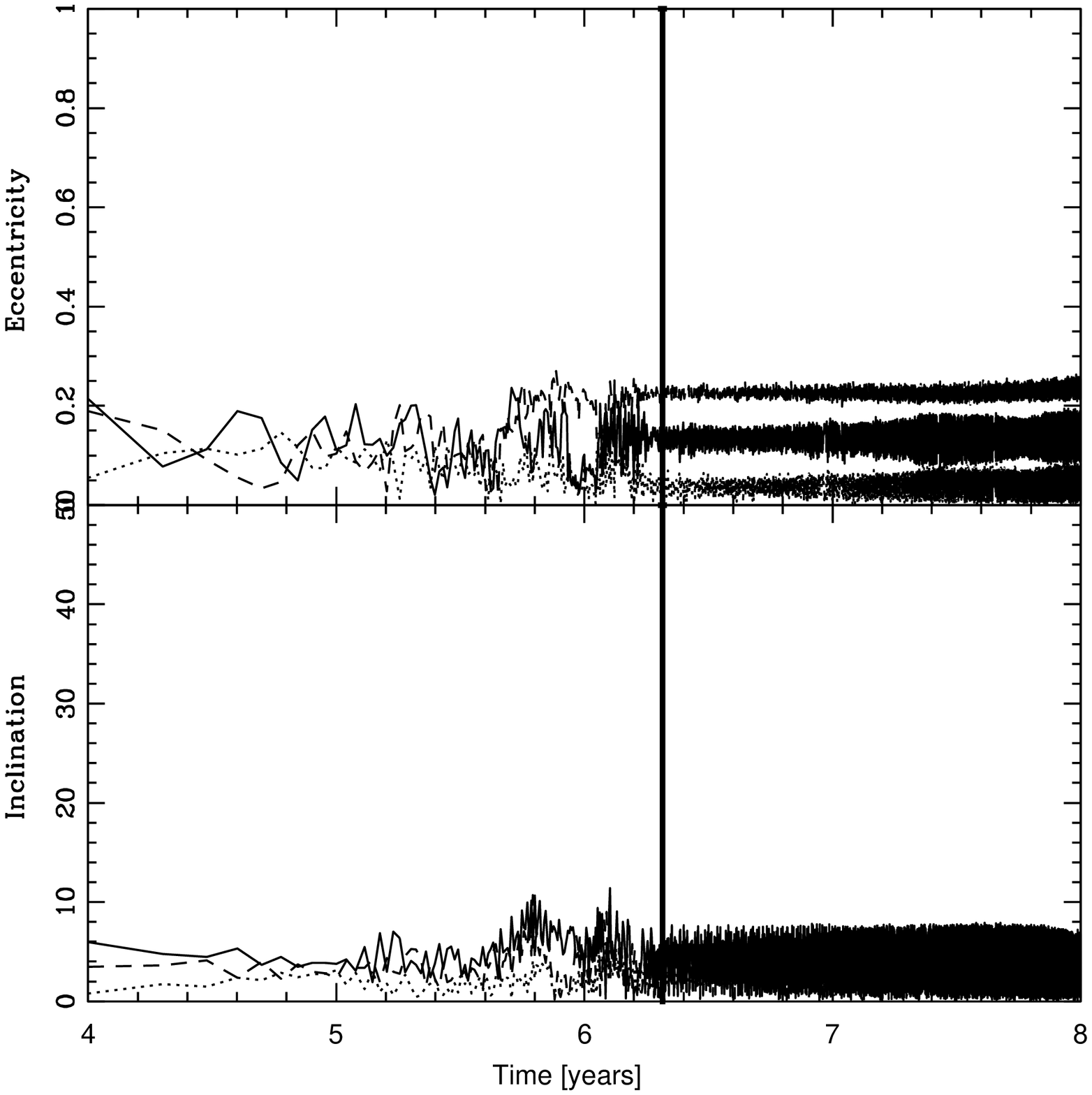}
\caption{Evolution of a three-planet system. Left: Semi-major axis
evolution of three planets. Colored contours show the surface mass density of a gas disk. 
The vertical truncation of the surface mass density corresponds to $\tau_{GD}$.
Right: Corresponding $e$ and $i$ evolution. The vertical line indicates $\tau_{GD}$. 
No instability occurs for $10^8$ yr. \label{fig4}}
\end{figure}

\begin{figure}
\plottwo{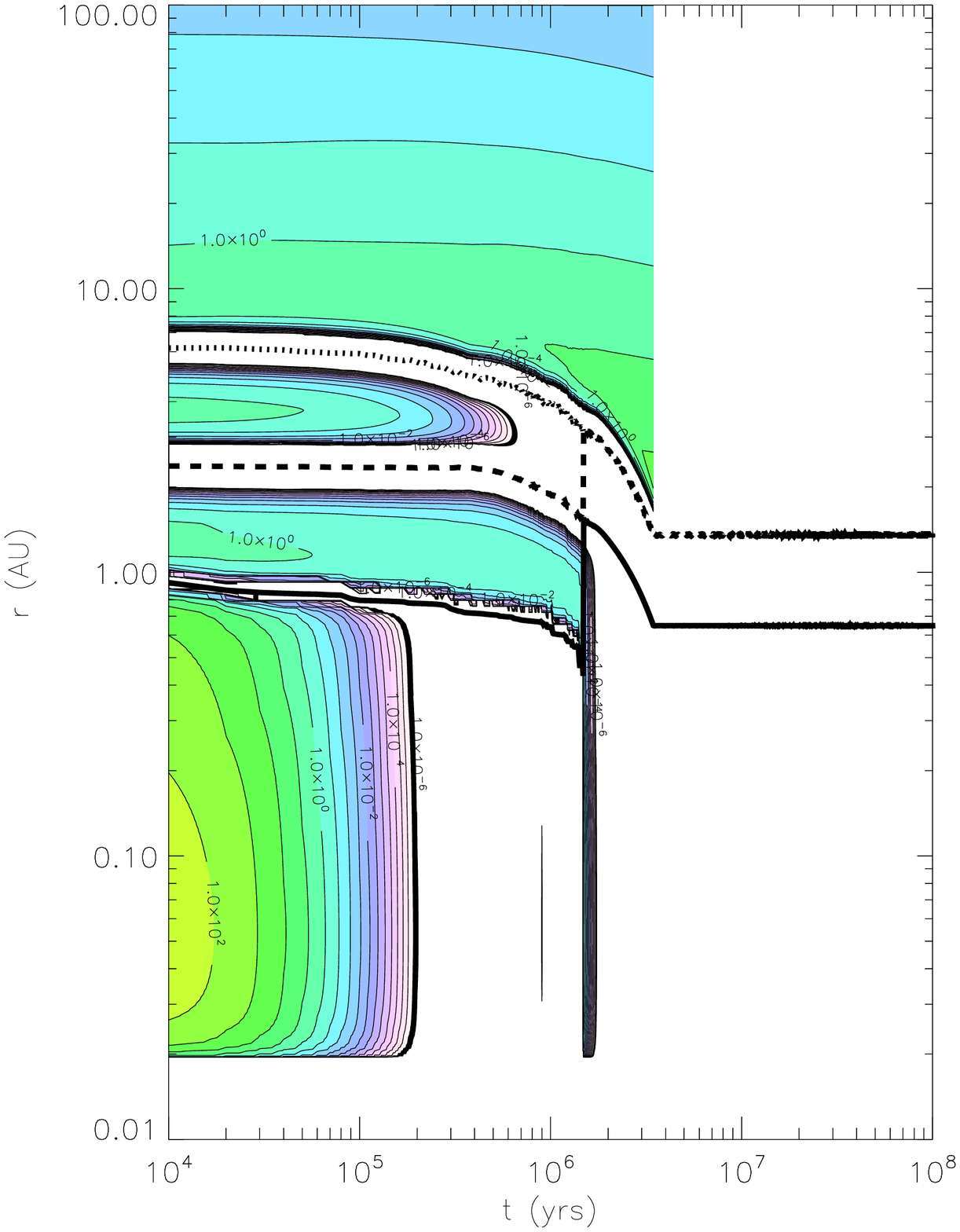}{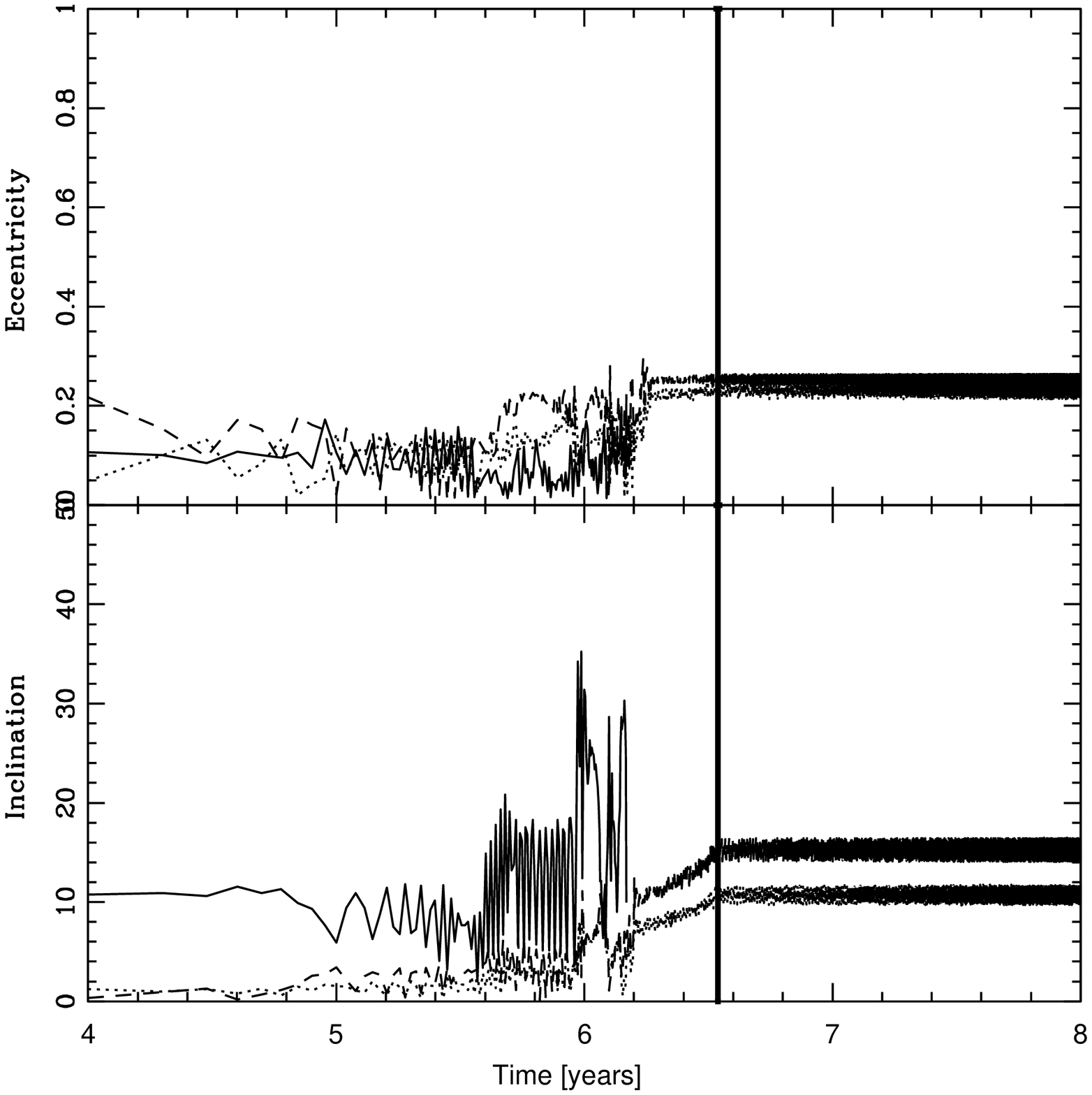}
\caption{Evolution of a three-planet system similar to Fig.~\ref{fig4}. Left: Semi-major axis
evolution of three planets. Right: Corresponding $e$ and $i$
evolution. Dynamical instability before $\tau_{GD}$ leads to an ejection of a planet out of the system. \label{fig5}}
\end{figure}

\begin{figure}
\plottwo{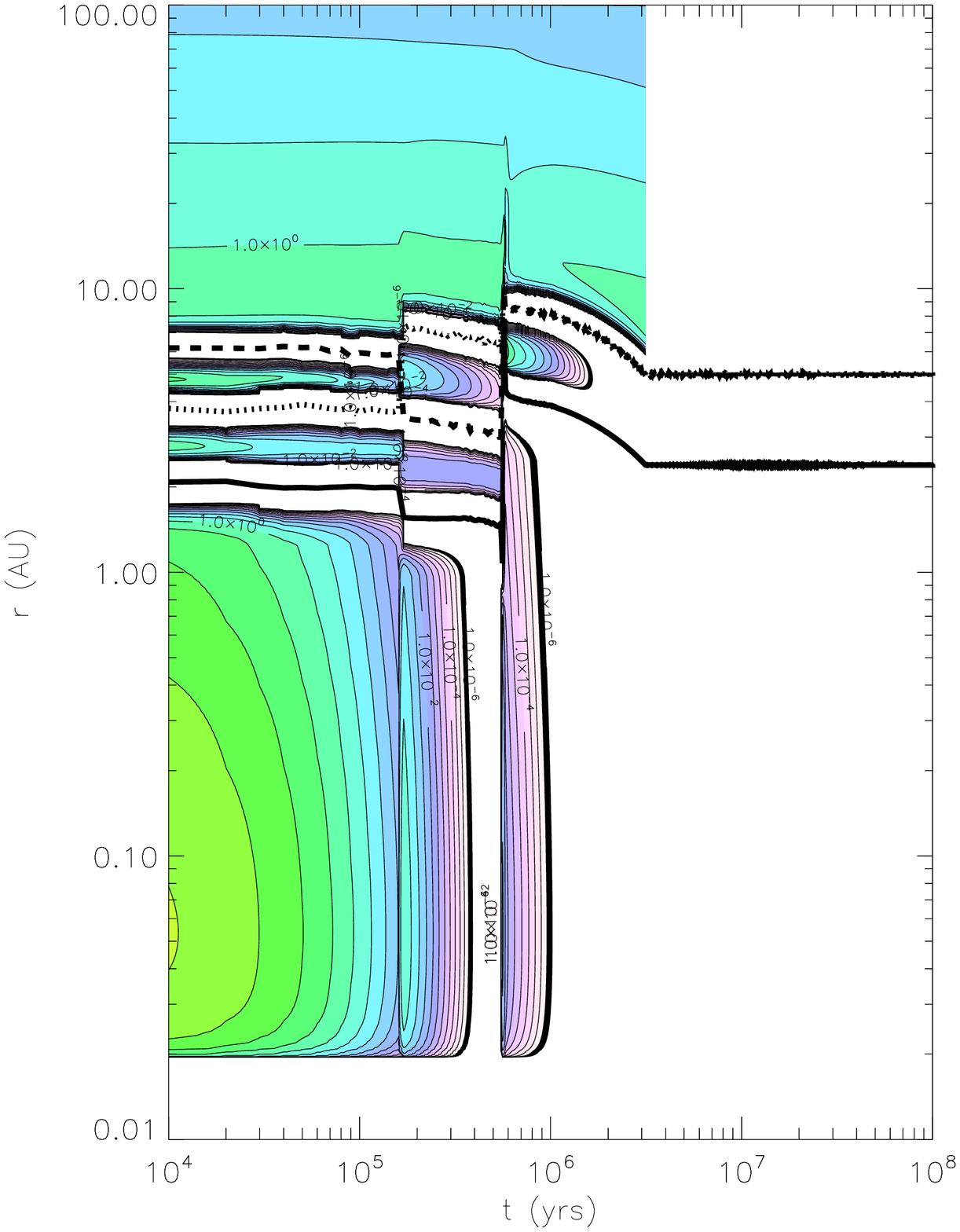}{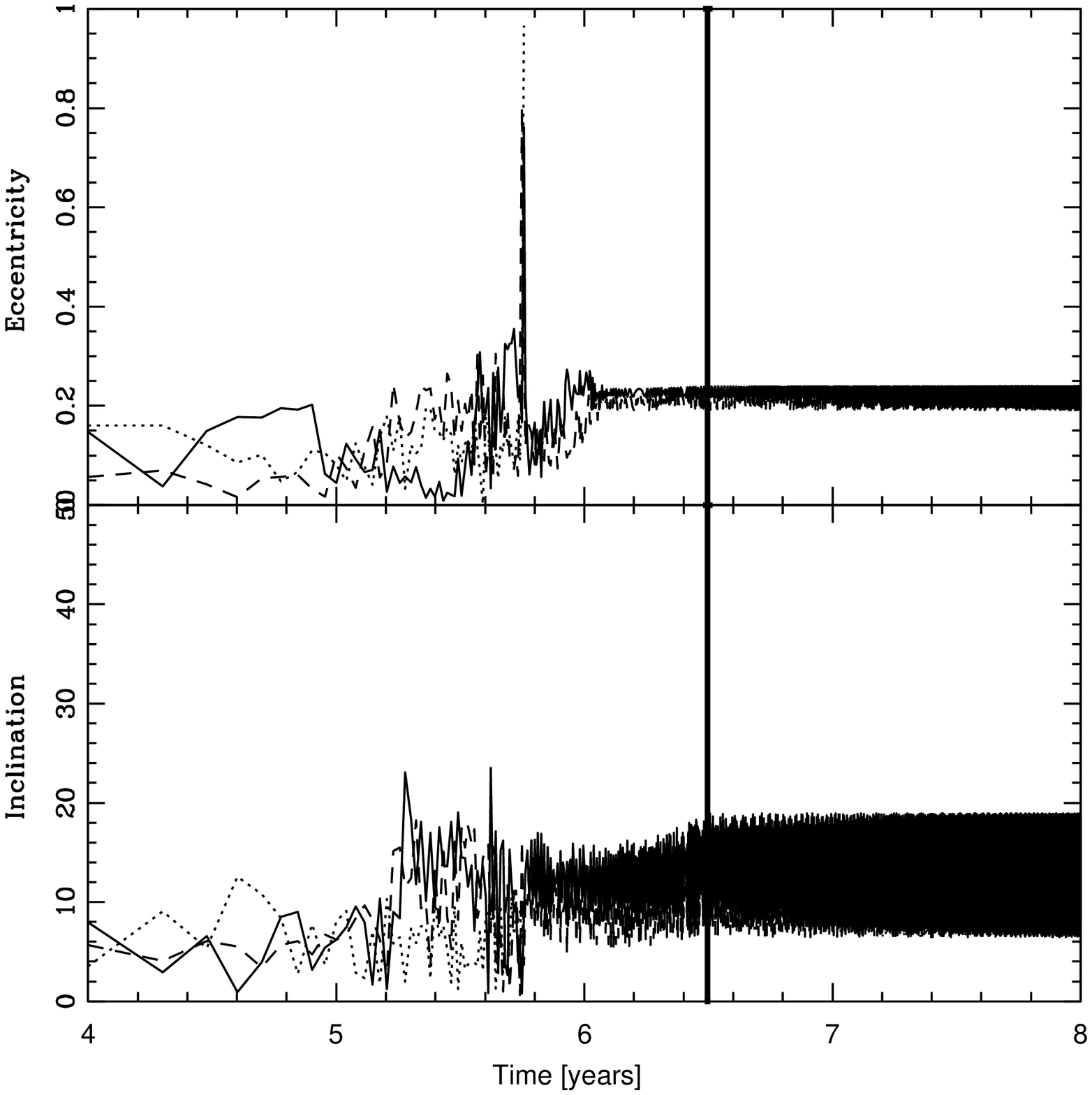}
\caption{Evolution of a three-planet system similar to Fig.~\ref{fig4}. Left: Semi-major axis
evolution of three planets. Right: Corresponding $e$ and $i$
evolution. Dynamical instability before $\tau_{GD}$ leads to a collision of a planet with the central star. \label{fig6}}
\end{figure}

\begin{figure}
\plottwo{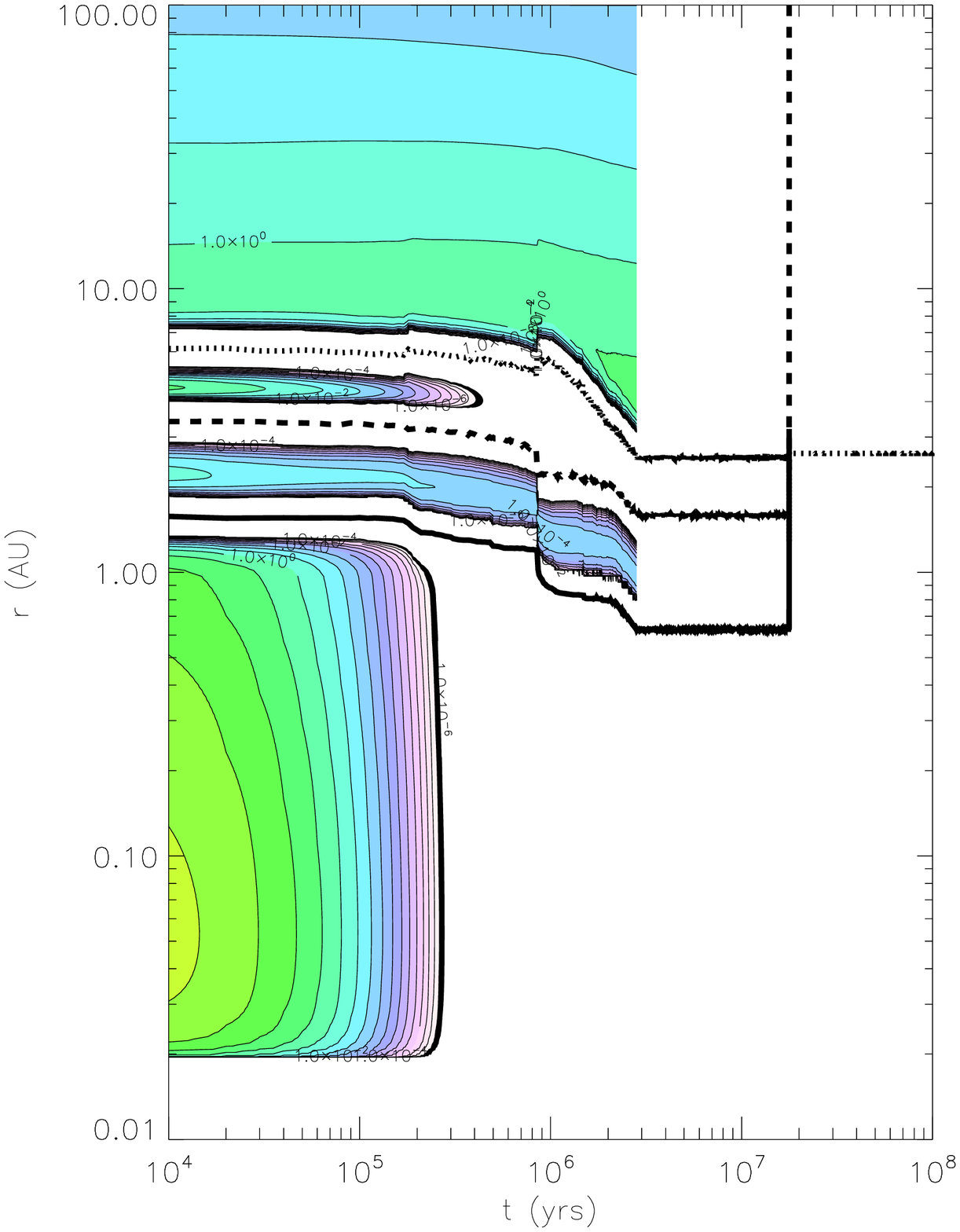}{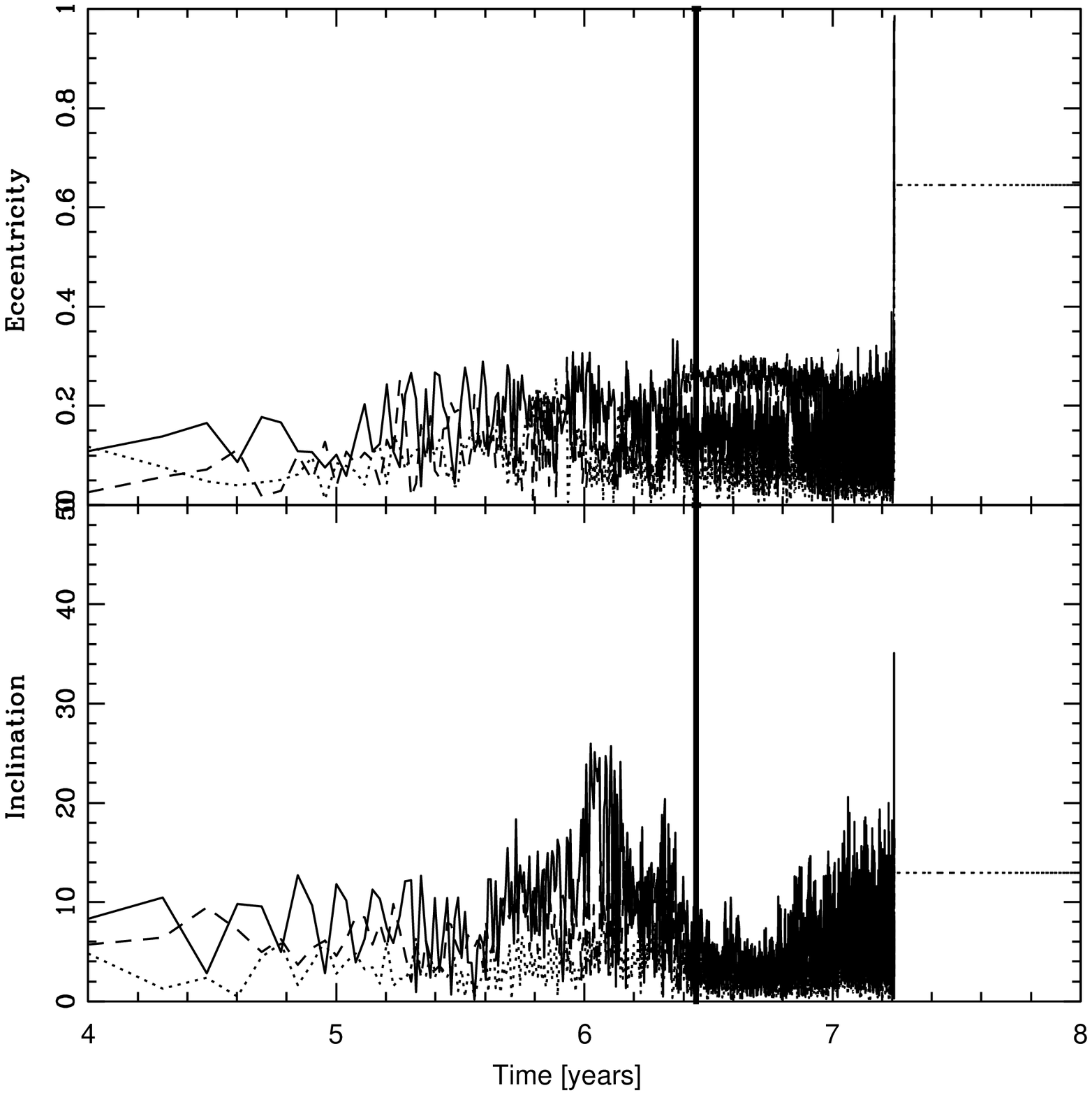}
\caption{Evolution of a three-planet system similar to Fig.~\ref{fig4}. Left: Semi-major axis
evolution of three planets. Right: Corresponding $e$ and $i$
evolution. Dynamical instability occurs long after the disk dissipation.
One planet is ejected, while another one collides with the central
star. \label{fig7}}
\end{figure}

\begin{figure}
\plottwo{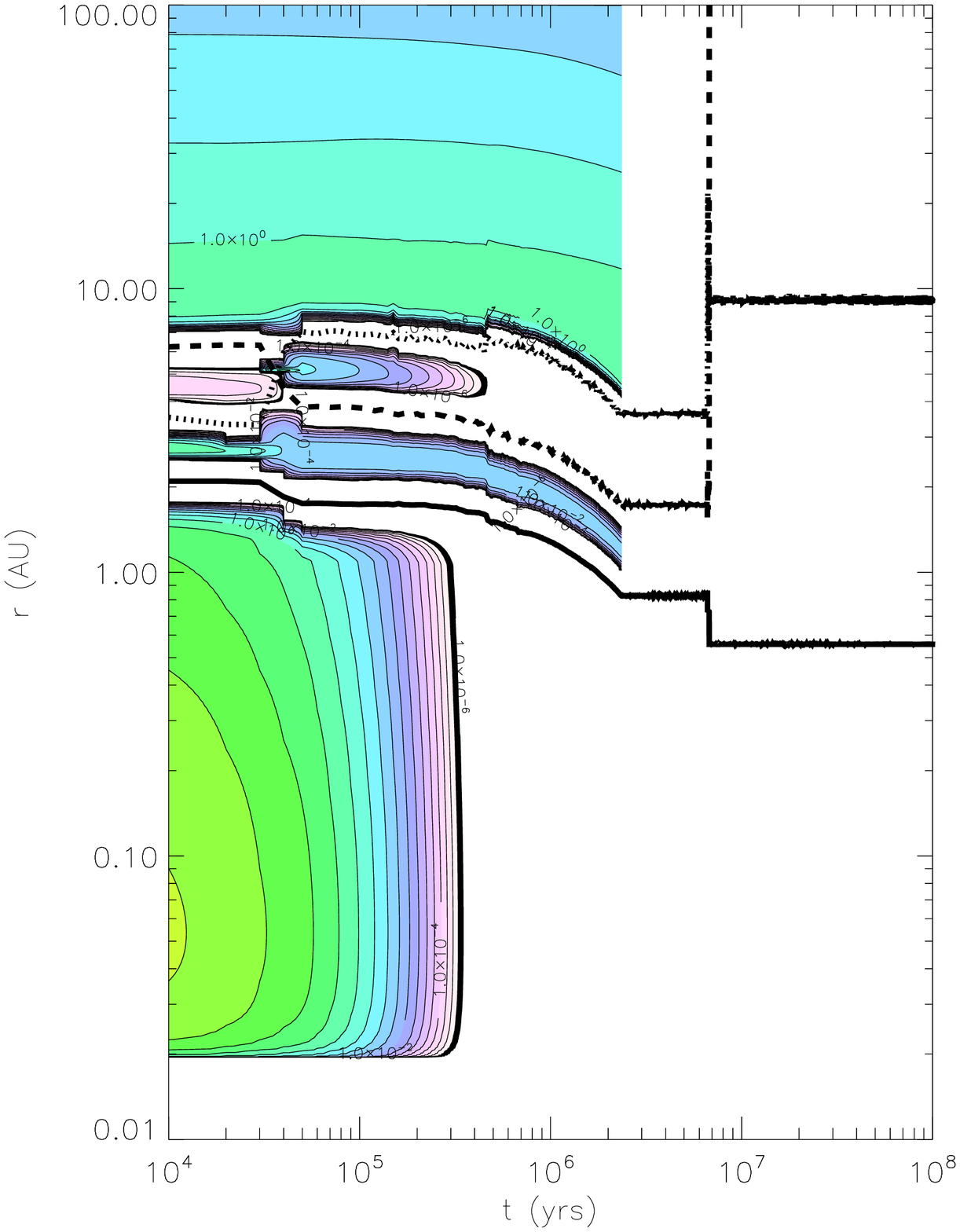}{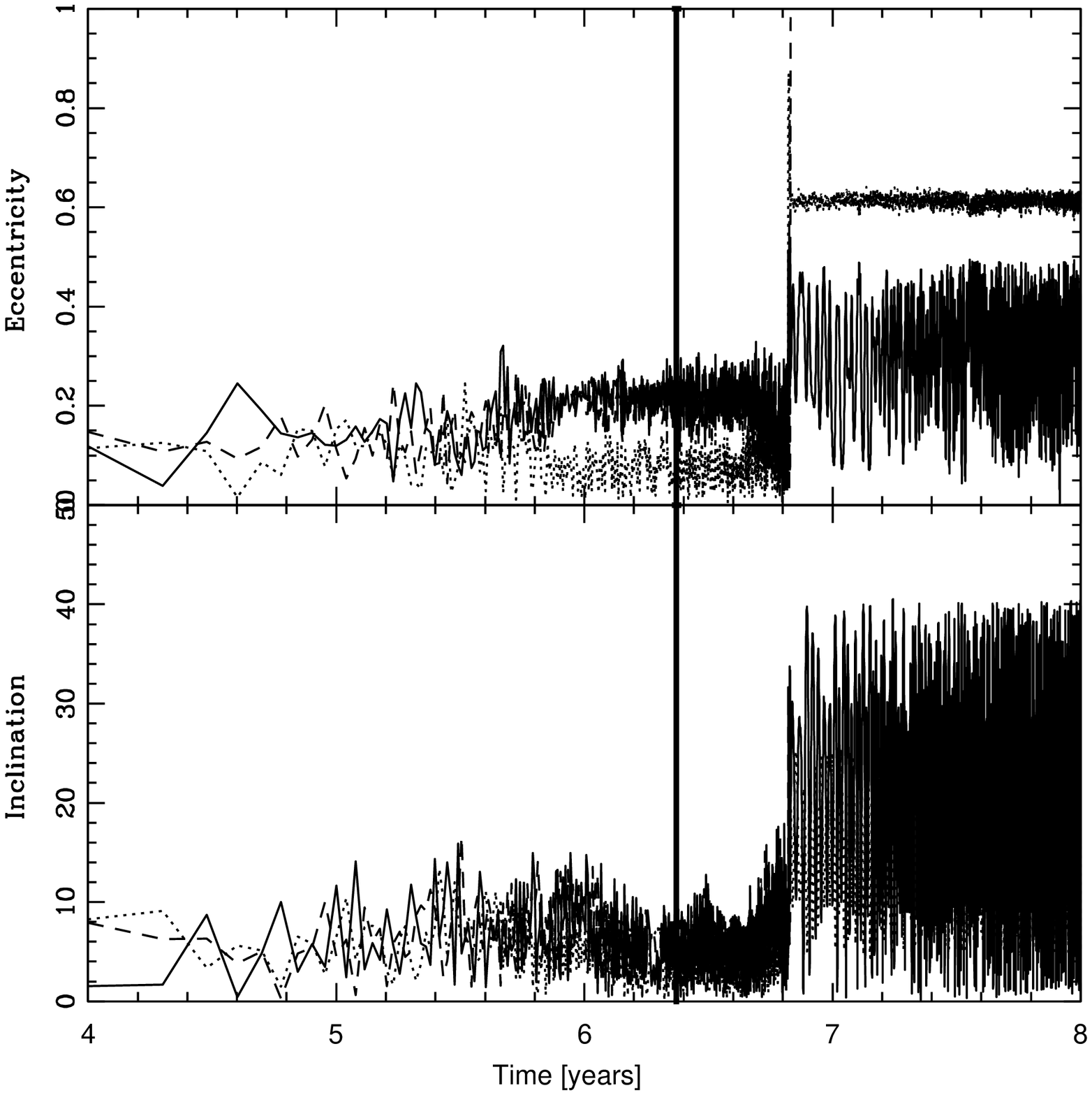}
\caption{Evolution of a three-planet system similar to Fig.~\ref{fig4}. Left: Semi-major axis
evolution of three planets. Right: Corresponding $e$ and $i$
evolution. Dynamical instability occurs after the disk dissipation. A
planet collides with the central star, leaving the other two planets
on eccentric, and inclined orbits. \label{fig8}}
\end{figure}

Fig.~\ref{fig4}-\ref{fig8} show some examples of evolution of
three-planet systems. Fig.~\ref{fig4} is a dynamically ``stable''
system, in which there are no ejections, mergers, nor collisions for
$100\,$Myr. Among the survived systems of our ``successful'' sets,
which have the K-S probability of the $a-e$ scattered plots of
$P>0.1$ ($t6d1$, $t7d1$, $t6d1cr$, $t7d1cr$, $t6d2$, $t7d2$,
$t7d2cr$, and $t8d2cr$), the fraction of these ``stable'' cases increases 
as the initial disk mass decreases, and spans over $\sim3-19\%$.
%

Fig.~\ref{fig5} and \ref{fig6} show the cases where dynamical
instability occurs while the gas disk is around. In the former
figure, gravitational interactions between planets lead to an ejection of a planet,
while in the latter, those lead to a collision with the central
star. In these cases, an ejection/collision does not result in a high
eccentricity of the remaining planets, probably due to the
eccentricity damping in the disk.
We find that about $40\%$ of the planetary systems in the successful sets 
experience either a collision or an ejection \emph{before} the gas disk dissipates.
%
%

Fig.~\ref{fig7} and \ref{fig8} show the cases where dynamical
instability occurs long after $\tau_{GD}$. In the former figure,
the strong interactions between planets lead to an ejection of one of the planets,
and a collision of another planet with the central star. The planet
left behind gains a large eccentricity. In the latter figure, such interactions lead to a collision of a planet with the
central star, again, leaving the other planets on eccentric, and
inclined orbits.
About $40\%$ of systems in the successful sets become dynamically unstable \emph{after} the disk dissipation.

The $a-e$ distribution for all the successful sets show the same trend as the representative case in Section 4.1 --- 
the observed $a-e$ distribution is poorly reproduced at $\tau_{GD}$, while at the end of the simulations (100\,Myr), 
the observed distribution is well reproduced.
Thus, although the rates of systems which become dynamically unstable are similar before and after $\tau_{GD}$, 
we find that the eccentricity distribution is largely determined after the disk dissipation, as the previous N-body simulations 
implicitly assumed.

%
%

\subsection{Mass Distribution}
We also studied the $a-M$ and $e-M$ distributions for these sets. We
find that the sets with the K-S probability of the $a-e$ scattered
plots being $P>0.1$ also have a high K-S probability for $a-M$ and
$e-M$ distributions (see Table~\ref{tab3}).

\begin{figure}
\plottwo{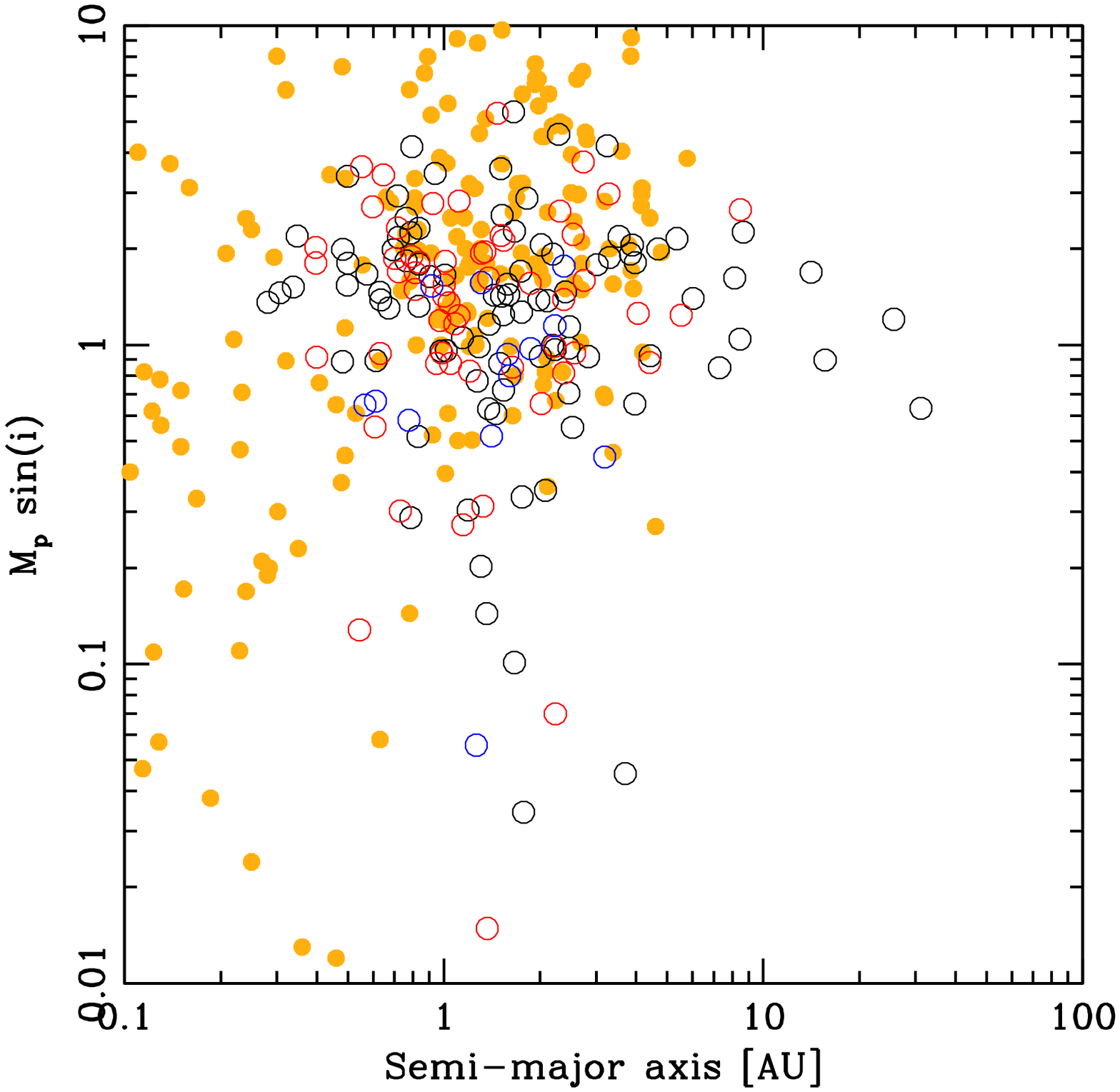}{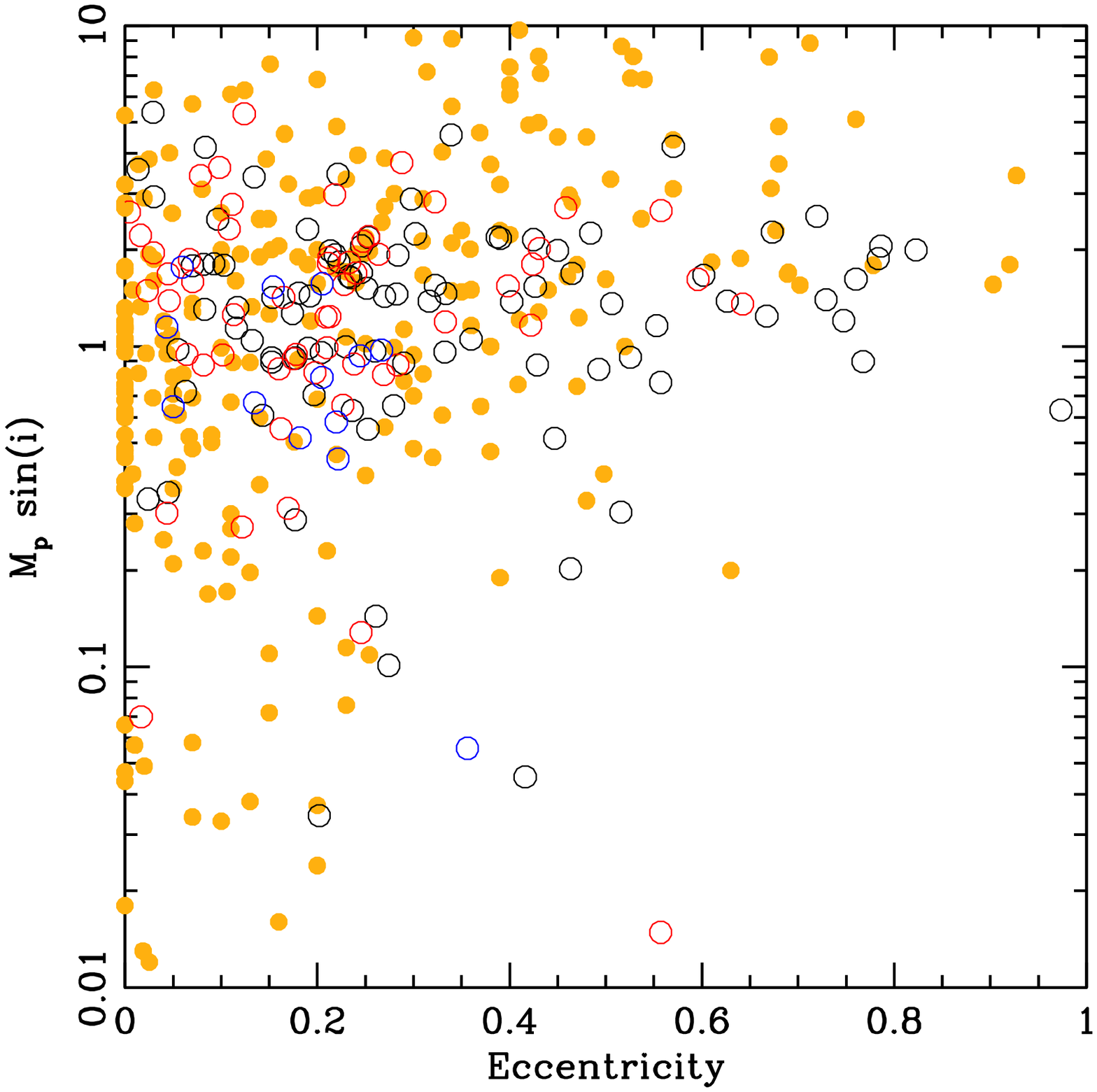}

\caption{Left: Semi-major axis and $M_p\sin i$ scattered plot for
$t7d1$. Right: The corresponding eccentricity and $M_p\sin i$
scattered plot. Here, $i$ is chosen randomly over $0-180\,$deg. \label{fig10}}
\end{figure}

In Fig.~\ref{fig10}, we plot $a$ and $e$ against $M_p\sin i$ for a
representative case of $t7d1$. The other cases ($t6d1$, $t6d1cr$,
$t7d1cr$, $t6d2$, $t7d2$, $t7d2cr$, and $t8d2cr$) look similar to
this. Here, we choose $i$ randomly from $0-180\,$deg for the
simulated planets. These plots mimic the expected $a-M$ and $e-M$
plots, assuming the simulated systems were observed, and the
planetary orbits were inclined randomly with respect to the plane of
the sky. From these plots, we expect that the future observations
will find lower mass planets ($M_p\sin i\lesssim 0.3 M_J$) on large
semi-major axis ($\gtrsim 1\,$AU).
%
%
%

\subsection{Mean Motion Resonances}
Although it's still too early to derive any statistical trend, some
of the extrasolar planetary systems are observed to be in mean motion resonances
(MMRs). Thus, it is interesting to investigate whether any of the
simulated systems are in such configurations.

At the end of the simulations, our ``successful'' cases, $t6d1$,
$t7d1$, $t6d1cr$, $t7d1cr$, $t6d2$, $t7d2$, $t7d2cr$, and $t8d2cr$ 
have $\sim20-70$ multi-planet systems.
For each of these systems, we estimate whether they are in a particular resonance by
using the following resonance variable \citep{Murray99}:
\begin{equation}
\varphi = j_1\lambda_{o}+j_2\lambda_{i}+j_3\varpi_{o}+j_4\varpi_{i}
\ ,
\end{equation}
where $\lambda$ and $\varpi$ are the mean longitude and longitude of
pericenter, respectively, and the subscripts $i$ and $o$ indicate
inner and outer planets.  Here, we focus on near coplanar cases, and
thus neglect the terms regarding the longitude of ascending node.
When planets are in $p+q:p$ MMR, we can define $(j_1,\, j_2,\,
j_3,\, j_4)=(p+q,\, -p,\, -q,\, 0)$, or $(p+q,\, -p,\, 0,\, -q)$.

We follow first- to third-order resonances (2:1, 3:2, 3:1, 5:3, 4:1,
5:2), as well as some higher order resonances (5:1, 7:3, 6:1, 7:2,
7:1, 8:1, 9:2, 9:1, 10:1).  
In Table~\ref{tab4}, we summarize the numbers and kinds of MMRs seen in our systems 
at the end of the simulations. We find that most of these systems get trapped in MMRs while 
a gas disk is around, either via migration, or planet--planet scattering. 
A recent N-body study of three-planet systems (with no gas disk) by \cite{Raymond08} 
showed that planet--planet scatterings can populate both low- and high-order MMRs. 
Our simulations confirm their result, and suggest that 
the combined effect of disk--planet and planet--planet interactions can generate planets in a variety of MMRs.

In our simulations, most multi-planet systems turn out to be in MMRs ($\sim 70-95\%$), while in \cite{Raymond08}, 
roughly $5-10\%$ of dynamically unstable systems ended up being in MMRs. 
This high rate of planets in MMRs seen in our simulations is clearly inconsistent with the observed systems. 
\cite{Adams08} showed that the stochastic forcing effects of turbulence tend to
pull planets out of MMRs. If this is the case, the number of planets
in MMRs may be much lower.
%
%

\section{Discussions and Conclusions}

We have studied the evolution of multiple-planet systems in a dissipating gas disk
by means of the hybrid code that combines the N-body symplectic
integrator SyMBA, and a one-dimensional gas disk evolution code. 
We simulate disk accretion, gap-opening by planets, and planet migration in a self-consistent manner, and 
also take account of the effects of eccentricity damping by disk--planet interactions as well as gas accretion by planets.
The main goal of this study is to investigate different plausible
scenarios and understand how various initial conditions affect the
final distributions of observable orbital properties. Specifically,
we have investigated the evolution of three-planet systems in a gas disk by utilizing a range of 
planetary and orbital properties (e.g.,~$M_p$, $a$, $e$, and $i$), as well as disk masses, 
and taking account of the absence/presence of saturation of corotation resonances.  

The initial conditions of our planetary systems are motivated by the standard planet formation theory.
We assume that giant planets are formed via core accretion, beyond the ice line, on initially nearly circular, and coplanar orbits (see Section 3).
Starting with three-planet systems that are initially fully embedded in gas disks, we have shown that the observed $a-e$ 
distribution can be well reproduced by such models (see Section 4).  

It is interesting to further compare the orbital properties of our simulations with current observations. 
Recently, \cite{Wright09} pointed out  possible differences in distributions of orbital parameters between 
single- and multiple-planet systems.  They found that the highly eccentric orbits ($e>0.6$) are predominantly 
associated with apparently single-planet systems.  The results of our simulations agree with this, and show 
that the high eccentricities ($e\gtrsim 0.5-0.6$) predominantly occur among single-planet systems at the end of the simulations.
This implies that currently observed apparently single-planet systems may have experienced strong dynamical instabilities in the past.

They also pointed out that planets with minimum planetary mass of $>1\,M_J$ have a broader eccentricity distribution, 
compared to less massive ones.
Although planetary masses in our simulations span only over $0.4-4\,M_J$, we do find a broader eccentricity distribution for planets with $>1\,M_J$.
Such a trend may be naturally explained as a result of strong gravitational interactions between planets, which tend to remove the 
less massive planet out of the system via an ejection, or a collision, leaving the more massive one on an eccentric orbit. 

Another striking difference \cite{Wright09} pointed out is the difference in semi-major axis distributions for single- and multiple-planet systems.  
The $a$ distribution of single-planet systems is characterized by the so-called 3-day peak and the jump in planetary abundance beyond 1 AU, while 
the corresponding distribution for multiple-planet systems is rather uniform.  
Such a difference may be well-explained if the evolution of single-planet systems has been dominated by planet--planet scatterings as opposed to 
planet migration, while that of multiple-planet systems has been dominated by planet migration.
Although we do not see such a difference in orbital distributions between single- and multiple-planet systems in our simulations, this may be partly because we don't have good numerical resolutions for $<0.2\,$AU. 
Interestingly, we find a group of planets with a range of $e$ that are located around $0.1\,$AU 
just before they are ejected out of the systems, or collides with the central star. 
Although our code does not include the tidal evolution directly, the separate calculations of tidal evolution of these systems indicate that they 
could be candidates of observed close-in, tidally-affected planets. 
\cite{Nagasawa08} pointed out that such close-in planets can be formed even without migration in gas disks. 
They studied the orbital evolution of three-planet systems with Jupiter mass by including the tidal circularization, 
and showed that Kozai mechanism combined with tidal orbital circularization (i.e.,~Kozai migration) can form 
these close-in planets for $\sim30\%$ cases of their simulations. 
 
Recent observations of Rossiter-McLaughlin effects started revealing an interesting subset of close-in exoplanets 
which have clearly misaligned orbits, or even retrograde ones \citep[e.g.,][]{Winn09,Narita09}.
Out of at least 17 transiting systems with the observed projected stellar obliquity, 8 have a clearly misaligned orbit ($>10\,$deg), 
among which 3 are on retrograde orbits.
We find that, from both simulations of three-planet systems with and without gas disks, 
the fractions of clearly inclined planets are$\sim15\%$ and $\sim65\%$, respectively, while 
the rate of planets on retrograde orbits is less than $1\%$ for both types of simulations.  
Thus, our results indicate that mechanisms other than planet--planet scatterings may be responsible 
for the observed high rate of close-in, retrograde planets.
Note, however, \cite{Nagasawa08} showed that inclinations could be more broadly distributed 
if close-in planets are formed via Kozai migration.
If this is the case, the inclination distribution of close-in planets may be significantly different from that of planets on further orbits.

Our investigations are affected by several significant
uncertainties.

First of all, the initial conditions for this kind of simulations
are highly uncertain. For example, we assume that nearly
fully-grown giant planets are initially embedded in a gas disk. In
reality, however, planets would start opening gaps as they grow, and
therefore they should not be fully embedded in a gas disk initially.
However, the planets in our simulations open gaps in a time on the
order of the orbital periods (i.e. less than several tens of years),
which are shorter than, or at most comparable to, both the
dynamical, and migration timescales. Therefore, we don't expect a
huge difference in the outcome due to this approximation.

Also, our choice of the initial planetary properties like
mass, and semimajor axis, although motivated by
the core accretion scenario, is rather arbitrary. 
To better approximate the initial conditions, we could have performed planet formation simulations as in \cite{Thommes08}.  
However, such simulations are computationally very expensive
for executing statistical studies.

Secondly, planet migration in our models do not
take account of the effects of corotation resonances. Since
corotation torques are sensitive to sharp gradients in the surface
mass density, they may have a significant effect on some of the planets
simulated here, which are massive enough to open a gap in the disk, but not 
massive enough to open a clean gap and saturate corotation resonances.
The so-called Type III migration due to the corotation torques tend to accelerate the inward planet migration
\citep{Masset03}, and thus we are likely to
underestimate planet migration rates for intermediate-mass
(Saturn-like) planets, which do not open a clean gap.
%

Regarding eccentricity evolution, we focus on the effects of
first-order resonances. Although this is a fair approximation for
planets with small eccentricities \citep{Goldreich80}, higher-order
resonances may become important if $e$ is excited to a large value
in the disk. Taking account of higher-order resonances,
\cite{Moorhead08} semi-analytically showed that the eccentricity
evolution timescales are decreased by a factor of a few, but the
overall trend of eccentricity evolution (e.g., damping or driving of
$e$) does not change. Furthermore, they suggested that eccentricity
decreases for $0.1\lesssim e\lesssim 0.5$ as long as corotation
resonances are unsaturated, while eccentricity can rapidly
increase when corotation resonances are fully saturated. 
On the other hand, current hydro simulations show
that disk-planet interactions excite eccentricity upto $\sim0.2$
\citep[e.g.][]{DAngelo06}. 
As already mentioned, our simulations don't take account of eccentricity excitation effect 
due to disk--planet interactions.  Thus, our simulations may underestimate the eccentricity excitation rate 
for massive planets, which open a clean gap, and have a sufficient eccentricity to saturate corotation resonances.
Clearly, this is an issue which requires a further investigation.

Finally, our gas disk is removed exponentially once the randomly
selected disk dissipation time is reached.  Although such a disk
removal is included to mimic the effect of photoevaporation, we did
not model its physics directly. This is because the photoevaporation
rate is difficult to estimate accurately, due to its sensitivity to
the stellar flux, which in turn depends on the disk accretion rate,
as well as the stellar environment \citep{Matsuyama03}.

Bearing these in mind, our simulations successfully reproduced the general trends of the observed properties, 
starting with the initial conditions expected from core accretion scenario.
We summarize our findings below.

\begin{enumerate}
\item Although the occurrence of dynamical instabilities before and after the disk dissipation is comparable, 
the $e$ distribution is largely determined by planet--planet
interactions after $\tau_{GD}$.

\item The $a$ distribution is largely determined by disk--planet
interactions.  To explain the current $a$ distribution, a disk mass 
has to be comparable to, or larger than, a planetary mass.

\item There may be an optimum disk mass to reproduce the observed $a-e$ distribution.


\item Dynamical instabilities in a gas disk which involve
ejections/collisions/mergers tend not to lead to large
eccentricities of remaining planets.

\item Initially nearly coplanar systems tend to have a higher merger rate between planets.

\item For the range of disk masses we use, we do not see a significant difference 
in outcomes by taking account of the saturation of corotation resonances, and hence the
reduction of $e$ damping.

\item We find that the combined effects of disk--planet and planet--planet interactions lead to 
both low- and high-order MMRs.  Our results also indicate that, without perturbing effects (e.g.,~turbulence), 
too many planets may be trapped in MMRs. 

\item Starting with relatively well-aligned, prograde orbits,  we find that planet--planet interactions 
are not efficient in producing the planets on retrograde orbits.

\end{enumerate}
%
%
%

\acknowledgements{This work was supported by NSF Grant AST-0507727
(to F.~A.~R.) and by a Spitzer Theory grant, as well as a grant from
Natural Sciences and Engineering Research Council of Canada (to
E.~W.~T.). We thank an anonymous referee for careful reading, and
helpful suggestions, which have improved this paper significantly.
S.~M. extends a sincere thank you to Ralph Pudritz for making
SHARCNET available for many of the simulations shown in this work.}

\bibliographystyle{apj}
\bibliography{REF}

\begin{deluxetable}{cccccccccc}
\tablecolumns{10}
\small
\tablewidth{0pt}
\tablecaption{Ejected, Collided, and Merged Planets \label{tab2}}
\tablehead{ & \multicolumn{3}{c}{Ejections}
            & \multicolumn{3}{c}{Collisions}
            & \multicolumn{3}{c}{Mergers} \\
           \colhead{Set No.}  &
           \colhead{Before $\tau_{GD}$}  & \colhead{After $\tau_{GD}$}  & \colhead{Total}  &
           \colhead{Before $\tau_{GD}$}  & \colhead{After $\tau_{GD}$}  & \colhead{Total}  &
           \colhead{Before $\tau_{GD}$}  & \colhead{After $\tau_{GD}$}  & \colhead{Total} }

\startdata
t5d1 & 110 (36.7) & 11 (23.4) & 121 & 21 (7) & 9 (19.1) & 30 & 58 (19.3) & 2 (4.26) & 60 \\
t6d1 & 62 (20.7) & 35 (23.6) & 97 & 6 (2) & 11 (7.43) & 17 & 42 (14) & 7 (4.73) & 49 \\
t7d1 & 45 (15.3) & 46 (21.6) & 91 & 1 (0.333) & 9 (4.23) & 10 & 25 (8.33) & 8 (3.76) & 33 \\
t8d1 & 35 (11.7) & 29 (13.6) & 64 & 0 (0) & 7 (3.27) & 7 & 15 (5) & 2 (0.935) & 17 \\
t9d1 & 25 (8.33) & 26 (10.2) & 51 & 0 (0) & 9 (3.53) & 9 & 12 (4) & 3 (1.18) & 15 \\
t5d1cr & 118 (39.3) & 16 (29.6) & 134 & 18 (6) & 3 (5.56) & 21 & 54 (18) & 4 (7.41) & 58 \\
t6d1cr & 83 (27.7) & 31 (27.9) & 114 & 11 (3.67) & 14 (12.6) & 25 & 41 (13.7) & 8 (7.21) & 49 \\
t7d1cr & 39 (13) & 46 (20.8) & 85 & 1 (0.333) & 10 (4.52) & 11 & 19 (6.33) & 9 (4.07) & 28 \\
t8d1cr & 47 (15.7) & 27 (12.1) & 74 & 0 (0) & 2 (0.893) & 2 & 14 (4.67) & 3 (1.34) & 17 \\
t9d1cr & 31 (10.3) & 36 (14.5) & 67 & 1 (0.333) & 2 (0.806) & 3 & 11 (3.67) & 3 (1.21)& 14 \\
t5d2 & 99 (33) & 14 (30.4) & 113 & 19 (6.33) & 3 (6.52) & 22 & 87 (29) & 3 (6.52) & 90 \\
t6d2 & 61 (20.3) & 31 (23.5) & 92 & 8 (2.67) & 5 (3.79) & 13 & 61 (20.3) & 3 (2.27) & 64 \\
t7d2 & 45 (15.3) & 29 (16.2) & 74 & 4 (1.33) & 5 (2.79) & 9 & 44 (14.7) & 7 (3.91) & 51 \\
t8d2 & 33 (11) & 19 (8.88) & 52 & 2 (0.667) & 5 (2.34) & 7 & 37 (12.3) & 4 (1.87) & 41 \\
t9d2 & 25 (8.33)  & 36 (16) & 61 & 0 (0) & 2 (0.889) & 2 & 41 (13.7) & 6 (2.67) & 47 \\
t5d2cr & 95 (31.7) & 12 (33.3) & 107 & 18 (6) & 1 (2.78) & 19 & 89 (29.7) & 1 (2.78) & 90 \\
t6d2cr & 53 (17.7) & 33 (23.6) & 86 & 11 (3.67) & 8 (5.71) & 21 & 62 (20.7) & 4 (2.86) & 66 \\
t7d2cr & 41 (13.7) & 30 (17.4) & 71 & 5 (1.67) & 2 (1.16) & 7 & 52 (17.3) & 6 (3.49) & 58 \\
t8d2cr & 28 (9.33) & 20 (10.1) & 48 & 3 (1) & 3 (1.51) & 6 & 51 (17) & 4 (2.01) & 55 \\
t9d2cr & 27 (9)  & 17 (7.83) & 44 & 0 (0) & 4 (1.84) & 4 & 43 (14.3) & 2 (0.922)& 45 \\
\enddata

\tablecomments{Numbers of planets which are ejected from the system,
collided with the central star, and merged with another planet,
before and after the gas dissipation time $\tau_{GD}$, as well as
throughout the simulations. Larger number of collisions before
$\tau_{GD}$ indicates more efficient planet migration, while the larger
number of ejections indicates higher occurrences of dynamical instability. The percentages
of planets in more than two-planet systems which are
ejected/collided/merged are shown inside the brackets.}

\end{deluxetable}
\clearpage
\begin{longtable}{ccccccccc}
\caption[tab3]{The 2D K-S test for $a$, $e$, and $M_p$ distributions \label{tab3}} \\

\hline \hline \\[-2ex]

\multicolumn{1}{c}{} &
\multicolumn{1}{c}{} & 
\multicolumn{2}{c}{$\tau_{GD}$ and $\tau_{fin}$ } & 
\multicolumn{2}{c}{$\tau_{GD}$ and Obs} & 
\multicolumn{2}{c}{$\tau_{fin}$ and Obs} \\
\multicolumn{1}{c}{Set No.} &
\multicolumn{1}{c}{} &
\multicolumn{1}{c}{D}  & 
\multicolumn{1}{c}{P}  & 
\multicolumn{1}{c}{D}  & 
\multicolumn{1}{c}{P}  & 
\multicolumn{1}{c}{D}  & 
\multicolumn{1}{c}{P} \\[0.5ex] \hline \\[-1.8ex]
\endfirsthead

\multicolumn{8}{c}{{\tablename} \thetable{} -- Continued} \\[0.5ex] 
\hline \hline \\[-2ex]
\multicolumn{1}{c}{} &
\multicolumn{1}{c}{} &
\multicolumn{2}{c}{$\tau_{GD}$ and $\tau_{fin}$ } &
\multicolumn{2}{c}{$\tau_{GD}$ and Obs} &
\multicolumn{2}{c}{$\tau_{fin}$ and Obs} \\
\multicolumn{1}{c}{Set No.} &
\multicolumn{1}{c}{} &
\multicolumn{1}{c}{D}  &
\multicolumn{1}{c}{P}  &
\multicolumn{1}{c}{D}  &
\multicolumn{1}{c}{P}  &
\multicolumn{1}{c}{D}  &
\multicolumn{1}{c}{P} \\[0.5ex] \hline \\[-1.8ex]
\endhead

%
%
\\[-1.8ex] \hline \hline \\
\multicolumn{9}{l}{
\tablecomments{
The results of the 2D K-S test for the combinations of
semimajor axis, eccentricity, and planetary masses.
We compare the simulated
results at $\tau_{fin}=100\,$Myr with the
observed planets between $0.2$ AU $\lesssim a\lesssim 6$ AU,
and $0.3 M_J \lesssim M_p \lesssim 4 M_J$.
The K-S statistic D, and the corresponding probability
P are shown for each comparison. We reject the null hypothesis for P
less than 0.1. 
Here, the null hypothesis is that the pair of samples
used in the test is drawn from the same distribution.}
} 
\endlastfoot

t5d1 & a vs e & $1.20e-2$ & $1.000$ & $6.95e-2$ & $5.55e-2$ & $7.80e-2$ & $1.94e-2$ \\
t5d1 & m vs e & $1.20e-2$ & $1.000$ & $6.70e-2$ & $5.75e-2$ & $7.80e-2$ & $1.71e-2$ \\
t5d1 & m vs a & $1.20e-2$ & $0.9999$ & $6.85e-2$ & $5.19e-2$ & $7.80e-2$ & $1.92e-2$ \\
\\
t6d1 & a vs e & $0.197$ & $0$ & $0.190$ & $0$ & $3.20e-2$ & $0.842$ \\
t6d1 & m vs e & $0.197$ & $0$ & $0.216$ & $0$ & $4.40e-2$ & $0.440$ \\
t6d1 & m vs a & $0.197$ & $0$ & $0.209$ & $0$ & $4.15e-2$ & $0.528$ \\
\\
t7d1 & a vs e & $0.281$ & $0$ & $0.314$ & $0$ & $3.70e-2$ & $0.693$ \\
t7d1 & m vs e & $0.281$ & $0$ & $0.325$ & $0$ & $5.65e-2$ & $0.172$ \\
t7d1 & m vs a & $0.281$ & $0$ & $0.328$ & $0$ & $5.40e-2$ & $0.212$ \\
\\
t8d1 & a vs e & $0.272$ & $0$ & $0.327$ & $0$ & $7.30e-2$ & $4.03e-2$ \\
t8d1 & m vs e & $0.272$ & $0$ & $0.341$ & $0$ & $7.90e-2$ & $1.60e-2$ \\
t8d1 & m vs a & $0.271$ & $0$ & $0.332$ & $0$ & $7.35e-2$ & $3.12e-2$ \\
\\
t9d1 & a vs e & $0.293$ & $0$ & $0.392$ & $0$ & $0.110$ & $0$ \\
t9d1 & m vs e & $0.293$ & $0$ & $0.395$ & $0$ & $0.112$ & $0$ \\
t9d1 & m vs a & $0.293$ & $0$ & $0.395$ & $0$ & $0.109$ & $0$ \\
\\
t5d1cr & a vs e & $2.00e-2$ & $0.998$ & $5.75e-2$ & $0.175$ & $7.15e-2$ & $4.29e-2$ \\
t5d1cr & m vs e & $1.85e-2$ & $0.999$ & $5.75e-2$ & $0.146$ & $7.10e-2$ & $3.83e-2$ \\
t5d1cr & m vs a & $1.85e-2$ & $0.9995$ & $5.60e-2$ & $0.180$ & $7.10e-2$ & $4.54e-2$ \\
\\
t6d1cr & a vs e & $0.164$ & $0$ & $0.125$ & $0$ & $5.35e-2$ & $0.238$ \\
t6d1cr & m vs e & $0.164$ & $0$ & $0.147$ & $0$ & $5.05e-2$ & $0.278$ \\
t6d1cr & m vs a & $0.164$ & $0$ & $0.154$ & $0$ & $5.10e-2$ & $0.281$ \\
\\
t7d1cr & a vs e & $0.299$ & $0$ & $0.334$ & $0$ & $5.20e-2$ & $0.269$ \\
t7d1cr & m vs e & $0.299$ & $0$ & $0.346$ & $0$ & $6.60e-2$ & $6.96e-2$ \\
t7d1cr & m vs a & $0.298$ & $0$ & $0.350$ & $0$ & $5.90e-2$ & $0.138$ \\
\\
t8d1cr & a vs e & $0.275$ & $0$ & $0.336$ & $0$ & $7.40e-2$ & $3.67e-2$ \\
t8d1cr & m vs e & $0.276$ & $0$ & $0.341$ & $0$ & $8.30e-2$ & $1.01e-2$ \\
t8d1cr & m vs a & $0.275$ & $0$ & $0.350$ & $0$ & $8.35e-2$ & $0$ \\
\\
t9d1cr & a vs e & $0.290$ & $0$ & $0.376$ & $0$ & $9.70e-2$ & $0$ \\
t9d1cr & m vs e & $0.288$ & $0$ & $0.381$ & $0$ & $0.101$ & $0$ \\
t9d1cr & m vs a & $0.288$ & $0$ & $0.381$ & $0$ & $0.101$ & $0$ \\
\\
t5d2 & a vs e & $1.65e-2$ & $0.9999$ & $6.60e-2$ & $8.45e-2$ & $8.10e-2$ & $1.64e-2$ \\
t5d2 & m vs e & $1.55e-2$ & $0.9999$ & $6.55e-2$ & $6.80e-2$ & $8.05e-2$ & $1.22e-2$ \\
t5d2 & m vs a & $1.60e-2$ & $0.9999$ & $6.55e-2$ & $7.51e-2$ & $8.05e-2$ & $1.41e-2$ \\
\\
t6d2 & a vs e & $0.173$ & $0$ & $0.157$ & $0$ & $4.10e-2$ & $0.541$ \\
t6d2 & m vs e & $0.172$ & $0$ & $0.177$ & $0$ & $5.25e-2$ & $0.236$ \\
t6d2 & m vs a & $0.173$ & $0$ & $0.178$ & $0$ & $4.80e-2$ & $0.344$ \\
\\
t7d2 & a vs e & $0.221$ & $0$ & $0.242$ & $0$ & $3.85e-2$ & $0.643$ \\
t7d2 & m vs e & $0.220$ & $0$ & $0.260$ & $0$ & $5.30e-2$ & $0.225$ \\
t7d2 & m vs a & $0.221$ & $0$ & $0.261$ & $0$ & $4.90e-2$ & $0.313$ \\
\\
t8d2 & a vs e & $0.245$ & $0$ & $0.314$ & $0$ & $8.25e-2$ & $1.29e-2$ \\
t8d2 & m vs e & $0.244$ & $0$ & $0.324$ & $0$ & $9.60e-2$ & $0$ \\
t8d2 & m vs a & $0.245$ & $0$ & $0.323$ & $0$ & $8.60e-2$ & $0$ \\
\\
t9d2 & a vs e & $0.277$ & $0$ & $0.324$ & $0$ & $6.95e-2$ & $5.63e-2$ \\
t9d2 & m vs e & $0.277$ & $0$ & $0.329$ & $0$ & $7.05e-2$ & $4.46e-2$ \\
t9d2 & m vs a & $0.277$ & $0$ & $0.337$ & $0$ & $7.60e-2$ & $2.27e-2$ \\
\\
t5d2cr & a vs e & $1.35e-2$ & $0.9999$ & $7.70e-2$ & $2.62e-2$ & $8.70e-2$ & $0$ \\
t5d2cr & m vs e & $1.20e-2$ & $0.9999$ & $7.65e-2$ & $1.92e-2$ & $8.70e-2$ & $0$ \\
t5d2cr & m vs a & $1.15e-2$ & $0.9999$ & $7.60e-2$ & $2.32e-2$ & $8.70e-2$ & $0$ \\
\\
t6d2cr & a vs e & $0.227$ & $0$ & $0.153$ & $0$ & $9.20e-2$ & $0$ \\
t6d2cr & m vs e & $0.227$ & $0$ & $0.175$ & $0$ & $9.25e-2$ & $0$ \\
t6d2cr & m vs a & $0.227$ & $0$ & $0.168$ & $0$ & $9.20e-2$ & $0$ \\
\\
t7d2cr & a vs e & $0.215$ & $0$ & $0.224$ & $0$ & $2.90e-2$ & $0.915$ \\
t7d2cr & m vs e & $0.214$ & $0$ & $0.242$ & $0$ & $4.80e-2$ & $0.330$ \\
t7d2cr & m vs a & $0.214$ & $0$ & $0.239$ & $0$ & $4.05e-2$ & $0.554$ \\
\\
t8d2cr & a vs e & $0.229$ & $0$ & $0.263$ & $0$ & $5.35e-2$ & $0.242$ \\
t8d2cr & m vs e & $0.229$ & $0$ & $0.271$ & $0$ & $6.15e-2$ & $0.108$ \\
t8d2cr & m vs a & $0.228$ & $0$ & $0.276$ & $0$ & $6.10e-2$ & $0.112$ \\
\\
t9d2cr & a vs e & $0.236$ & $0$ & $0.300$ & $0$ & $7.85e-2$ & $2.20e-2$ \\
t9d2cr & m vs e & $0.236$ & $0$ & $0.304$ & $0$ & $8.30e-2$ & $1.00e-2$ \\
t9d2cr & m vs a & $0.236$ & $0$ & $0.310$ & $0$ & $8.35e-2$ & $0$ \\
%
%
\end{longtable}

\begin{deluxetable}{lccccccccccccc}

\tablecolumns{14}
\small
\tablewidth{0pt}
\tablecaption{Numbers of Planetary Systems in MMRs  \label{tab4}}

\tablehead{ \colhead{Set No.} & \colhead{$2:1$}& \colhead{$3:2$}  & \colhead{$3:1$} & 
                      \colhead{$4:1$}  & \colhead{$5:2$}  & \colhead{$5:1$}  & \colhead{$6:1$} & 
                      \colhead{$7:2$}  & \colhead{$7:1$}  & \colhead{$8:1$}  & \colhead{$9:2$} &
                      \colhead{$9:1$}  & \colhead{$10:1$}  }

\startdata
t6d1 & 6 & 0 & 16 & 5 & 4 & 1 & 2 & 2 & 0 & 0 & 0 & 0 & 0 \\
t7d1 & 23 & 1 & 22 & 7 & 5 & 1 & 1 & 1 & 1 & 1 & 0 & 0 & 0 \\
t6d1cr & 5 & 0 & 6 & 5 & 2 & 0 & 0 & 1 & 0 & 1 & 0 & 0 & 0 \\
t7d1cr & 23 & 0 & 17 & 7 & 7 & 3 & 5 & 2 & 2 & 0 & 0 & 0 & 1 \\
t6d2 & 10 & 0 & 17 & 7 & 4 & 0 & 1 & 0 & 0 & 0 & 0 & 0 & 0 \\
t7d2 & 14 & 2 & 19 & 11 & 6 & 0 & 2 & 1 & 1 & 1 & 1 & 0 & 1 \\
t7d2cr & 14 & 2 & 19 & 11 & 6 & 0 & 2 & 1 & 1 & 1 & 1 &  0 & 1 \\
t8d2cr & 21 & 0 & 28 & 7 & 13 & 6 & 4 & 2 & 0 & 0 & 0 & 1  & 0 \\
\enddata

\tablecomments{The numbers of systems which have each kind of MMRs.  
The shown sets are successful cases which have $P>0.1$ for the K-S test of the observed and simulated $a-e$ scattered plot. 
Most of these planets are trapped in MMRs while a gas disk is around.}
\end{deluxetable}
%
%
%

\end{document}